\documentclass[%
 reprint,
 amsmath,amssymb,
 aps,
]{revtex4-1}

\usepackage{graphicx}
\usepackage{dcolumn}
\usepackage{bm}

\usepackage{graphicx}
\usepackage{dcolumn}
\usepackage{bm}
\usepackage{color}	
\usepackage{hyperref}
\usepackage{amsmath}
\begin{document}

\title{Rheo-acoustic gels: Tuning mechanical and flow properties\\of colloidal gels with ultrasonic vibrations}

\author{Thomas Gibaud}
\email{Corresponding author. thomas.gibaud@ens-lyon.fr}
\affiliation{Univ Lyon, Ens de Lyon, Univ Claude Bernard, CNRS, Laboratoire de Physique, 69342 Lyon, France}

\author{No\'emie Dag\`es}
\affiliation{Univ Lyon, Ens de Lyon, Univ Claude Bernard, CNRS, Laboratoire de Physique, 69342 Lyon, France}

\author{Pierre Lidon}
\affiliation{Univ Lyon, Ens de Lyon, Univ Claude Bernard, CNRS, Laboratoire de Physique, 69342 Lyon, France}

\author{Guillaume Jung}
\affiliation{Univ Lyon, Ens de Lyon, Univ Claude Bernard, CNRS, Laboratoire de Physique, 69342 Lyon, France}

\author{L. Christian Ahour\'e}
\affiliation{Univ Lyon, Ens de Lyon, Univ Claude Bernard, CNRS, Laboratoire de Physique, 69342 Lyon, France}

\author{Michael Sztucki}
\affiliation{ESRF -- The European Synchrotron, 38043 Grenoble, France}

\author{Arnaud Poulesquen}
\affiliation{CEA, DEN, Univ. Montpellier, DE2D, SEAD, LCBC, 30207 Bagnols-sur-C\`eze, France}

\author{Nicolas Hengl}
\affiliation{Univ. Grenoble Alpes, CNRS, Grenoble INP, LRP, 38000 Grenoble, France}

\author{Fr\'ed\'eric Pignon}
\affiliation{Univ. Grenoble Alpes, CNRS, Grenoble INP, LRP, 38000 Grenoble, France}

\author{S\'ebastien Manneville}
\affiliation{Univ Lyon, Ens de Lyon, Univ Claude Bernard, CNRS, Laboratoire de Physique, F-69342 Lyon, France}
\affiliation{MultiScale Material Science for Energy and Environment,
UMI 3466, CNRS-MIT, 77 Massachusetts Avenue, Cambridge, Massachusetts 02139, USA}

\date{\today}

\begin{abstract}
Colloidal gels, where nanoscale particles aggregate into an elastic yet fragile network, are at the heart of materials that combine specific optical, electrical and mechanical properties. Tailoring the viscoelastic features of colloidal gels in real-time thanks to an external stimulus currently appears as a major challenge in the design of ``smart'' soft materials. Here we introduce ``rheo-acoustic'' gels, a class of materials that are sensitive to ultrasonic vibrations. By using a combination of rheological and structural characterization, we evidence and quantify a strong softening in three widely different colloidal gels submitted to ultrasonic vibrations (with submicron amplitude and frequency 20--500~kHz). This softening is attributed to micron-sized cracks within the gel network that may or may not fully heal once vibrations are turned off depending on the acoustic intensity. Ultrasonic vibrations are further shown to dramatically decrease the gel yield stress and accelerate shear-induced fluidization. Ultrasound-assisted fluidization dynamics appear to be governed by an effective temperature that depends on the acoustic intensity. Our work opens the way to a full control of elastic and flow properties by ultrasonic vibrations as well as to future theoretical and numerical modeling of such rheo-acoustic gels.
\end{abstract}

\pacs{Valid PACS appear here}

\maketitle

\section{Introduction}

Colloidal gels constitute a class of soft materials with a huge spectrum of applications, ranging from paints, oil extraction and construction to pharmaceuticals and food products \cite{Larson:1999,Mewis:2012}. These gels are typically formed from the aggregation of attractive nanoscale particles that arrange into a space-spanning network which strength provides the system with elastic properties at rest. The particle network is, however, fragile enough that rather weak external forces disrupt the gel structure and induce flow above a critical yield strain or stress \cite{Bonn:2017}. Such unique mechanical and flow properties are key to applications that require an interplay between solid and fluid behaviour, including cement placement, ink-jet printing or flow-cell batteries. In this context, decades of research have investigated the influence of physico-chemical parameters such as temperature, pH and ionic strength on the aggregation of attractive particles into colloidal gels \cite{Cipelletti:2000,Romer:2000,Trappe:2001,Cardinaux:2007,Zaccarelli:2007,Lu:2008}. While electrostatic repulsion, van der Waals attraction and steric interaction, combined in the well-known DLVO interparticle potential, generically lead to the formation of fractal networks through diffusion- or reaction-limited cluster aggregation, additional chemical effects such as hydration layers or particle bridging generate a wealth of more complex gel microstructures \cite{Israelachvili:2011}. In the quest of smart, responsive materials, tuning the gel architecture in real-time and reversibly is a key challenge. However, 
playing on the above physico-chemical parameters turns out to be difficult, if not impossible when the chemistry of the material is strongly constrained. Electro- or magneto-rheological materials allow for such tuning through the use of electrical and magnetic fields \cite{Hao:2002,Kogan:2009,Swan:2012,Helal:2016a}. Yet, applications are obviously restricted to materials with specific compositions \cite{Coulter:1993,Tao:2001,Wereley:2013}.

The exquisite sensitivity of colloidal gels to external mechanical perturbations provides an alternative route to interact with their microstructure. The effects of steady or low-frequency oscillatory flows on the gel network are commonly probed through rheological measurements \cite{Mewis:2012}. Besides viscoelasticity and aging dynamics at rest \cite{Zaccarelli:2007}, the signature of the solidlike nature of colloidal gel networks is the existence of a yield stress: when the applied shear stress is large enough to break interparticle bonds, the gel network gets fragmented into a dispersion of aggregates. The yield stress typically shows a power-law increase with the particle volume fraction \cite{Buscall:1988,Chen:1991} that has been interpreted in the framework of fractal networks \cite{Potanin:1996,Shih:1999}. Beyond yielding, the dispersion flows and further shows shear-thinning as the typical aggregate size decreases with the shear rate \cite{Wessel:1992}. Superposition rheology, which consists in applying small-amplitude oscillatory shear on top of a steady shear flow, in a direction either parallel or orthogonal to the main shear flow \cite{Vermant:1998,Dhont:2001}, provides an interesting way to probe the viscoelastic properties of colloidal gels under nonlinear deformation \cite{Colombo:2017,Sung:2018}. Furthermore, to get microscopic insight into yielding, rheological measurements have been combined to structural characterization \cite{Poon:1997,Vermant:2005}. For instance, confocal microscopy has allowed for direct visualization of shear-induced breakup and relaxation after shear cessation \cite{Bonn:2009,Koumakis:2015}, for tracking of the erosion of rigid clusters by shear \cite{Hsiao:2012} and for detailed investigation of aggregate morphologies in two dimensional gels \cite{Masschaele:2011}. Small-angle light and x-ray scattering has been extensively used to study shear-induced anisotropy and restructuring, e.g., in silica gels \cite{Verduin:1996,Varadan:2001}, polystyrene spheres \cite{Mohraz:2005} and clay suspensions \cite{Pignon:1997a}. Small-angle neutron scattering coupled to rheology \cite{Eberle:2012} was also instrumental in revealing the shear-induced microstructure of colloidal gels both under steady shear \cite{Eberle:2014,Hipp:2019} and under oscillatory flow \cite{Kim:2014}. Finally, recent advances in numerical simulation schemes for colloidal particles under shear, including Brownian dynamics or Stokesian dynamics \cite{Zia:2010,Swan:2010,Seto:2012}, have allowed fruitful comparisons with experiments on shear-induced structuring of attractive particulate systems \cite{Koumakis:2015,Landrum:2016,Moghimi:2017,Varga:2019}.

All of this previous work resulted not only in a better fundamental understanding of the complex interplay between flow and gel structure but also in the optimization of both manufacturing processes and final product properties of an overwhelming number of colloid-based materials \cite{Bergna:2005,Mezzenga:2005,Guvendiren:2012}. In practice, however, applying a mechanical stress in order to fluidize the gel at will and {\it in situ} remains challenging as it involves moving parts such as pumps, motors or other rotating tools that may not be compatible with the application. Here, we introduce ultrasonic vibrations, i.e. high-frequency mechanical vibrations, as a way to control the elastic modulus and/or the yield stress of a colloidal gel. High-power ultrasound has long been known to disrupt particulate aggregates with applications to resuspension, filtration and cleaning \cite{Kyllonen:2005,Hengl:2014}, optimization of food products \cite{Knorr:2004,Awad:2012,Chandrapala:2012} and drug delivery \cite{Mitragotri:2005,Huebsch:2014}. Conversely, intense acoustic waves may induce gelation of some organic compounds and colloidal systems \cite{Cravotto:2009,Bardelang:2009}. Yet, in spite of such empirical know-how, no systematic study has been performed on the effect of ultrasonic vibrations on colloidal gels and it remains unclear how to distinguish between physico-chemical effects (e.g. chemical reactions triggered by cavitation and local temperature rise) \cite{Suslick:1999} and mechanical effects (e.g. material deformation due to acoustic radiation pressure or large-scale flow due to acoustic streaming) \cite{Hamilton:1998,Dalecki:2004}. Moreover, besides applications to material design, the interplay between ultrasound and the complex structure of colloidal gels raises a fundamental challenge due to the large span of time- and length-scales involved in the problem. 

In the present work, we probe the effects of ultrasonic vibrations on colloidal gels thanks to two different experimental setups designed to measure the gel mechanical and structural properties under ultrasound. Vibrations are generated by piezo-electric transducers driven at frequencies ranging from 20 to 500~kHz in direct contact with the gels. The corresponding acoustic wavelength ranges from 3~mm to several centimeters, which is always larger than the thickness of the samples under study. Therefore, in a first approximation, the gels should be considered as vibrated homogeneously at a high frequency rather than submitted to propagating ultrasonic waves or to a stationary ultrasonic field with pressure nodes and antinodes which position would depend on the boundary conditions. Our approach differs from the superposition techniques mentioned above in that we probe the low-frequency rheological response to vibrations with frequencies that are inaccessible to standard rheometry. In other words, there is a very large separation between the time scale at which the system is forced and that at which it is probed.

We focus on three different colloidal gels respectively composed of nanoparticles of calcite, silica and carbon black dispersed at low volume fractions in a background solvent. As described in more details in Appendix~\ref{app:gels}, all systems aggregate into elastic space-spanning networks thanks to attractive interparticle forces. At rest, once gelation is complete, all gels are solid and display an elastic modulus $G'$ and a yield stress $\sigma_\text{y}$, the stress above which the gel starts to flow. These gels are representative of the colloidal gels commonly encountered in soft matter, featuring elastic moduli $G'$ from about 100~Pa to 10~kPa and a wide range of yielding behaviours from ``simple'' yielding with little time dependence for the calcite gel to ``delayed yielding'' for the carbon black gel \cite{Gibaud:2010,Bonn:2017}.

The paper is organized as follows. In Sect.~\ref{sec:softening}, we first show that the elastic modulus of our colloidal gels decreases dramatically under ultrasonic vibrations. Such ultrasound-induced softening defines ``rheo-acoustic'' gels, i.e. gels which elastic properties are strongly dependent on ultrasonic vibrations. In Sect.~\ref{sec:saxs}, we couple x-ray scattering to ultrasonic vibrations and provide evidence for structural modifications hinting at ultrasound-induced micron-sized cracks within the gel network. Section~\ref{sec:flow} further explores the way ultrasonic vibrations may modify the flow and yielding properties of colloidal gels. Ultrasonic vibrations are shown to facilitate yielding and a tentative interpretation is proposed in terms of an effective temperature. Section~\ref{sec:conclu} discusses the various original questions opened by our experiments with emphasis on the need for a better microscopic understanding of the coupling between ultrasonic vibrations and the gel network. We close this article by discussing the implications of our results for future theoretical and numerical studies.

\section{Gel softening under high-power ultrasonic vibrations}
\label{sec:softening}

\paragraph*{Experimental setup.}
The mechanical effects of ultrasonic vibrations on colloidal gels are investigated thanks to a rotational rheometer (Anton Paar MCR 301) equipped with a parallel-plate geometry where the upper plate is made of sandblasted Plexiglas. The bottom plate is constituted of a piezoelectric transducer working either at frequency $f=45$~kHz (Sofranel, diameter of 29~mm) or 500~kHz (Imasonic, diameter of 50~mm). The piezoelectric transducer is fed with an oscillating voltage of amplitude up to 90~V through a broadband power amplifier (Amplifier Research 75A250A) driven by a function generator (Agilent 33522A). A thermocouple (National Instruments USB-TC01) monitors the temperature at the surface of the transducer. A solvent trap covers the geometry in order to minimize evaporation. The gap width is set to 1~mm for all experiments except in Fig.~\ref{fig:modulus_geometry} in Appendix~\ref{app:geometry} where the influence of the gap width is tested.

 \begin{figure}[ht!]
 	\centering
 	\includegraphics[width=0.9\columnwidth]{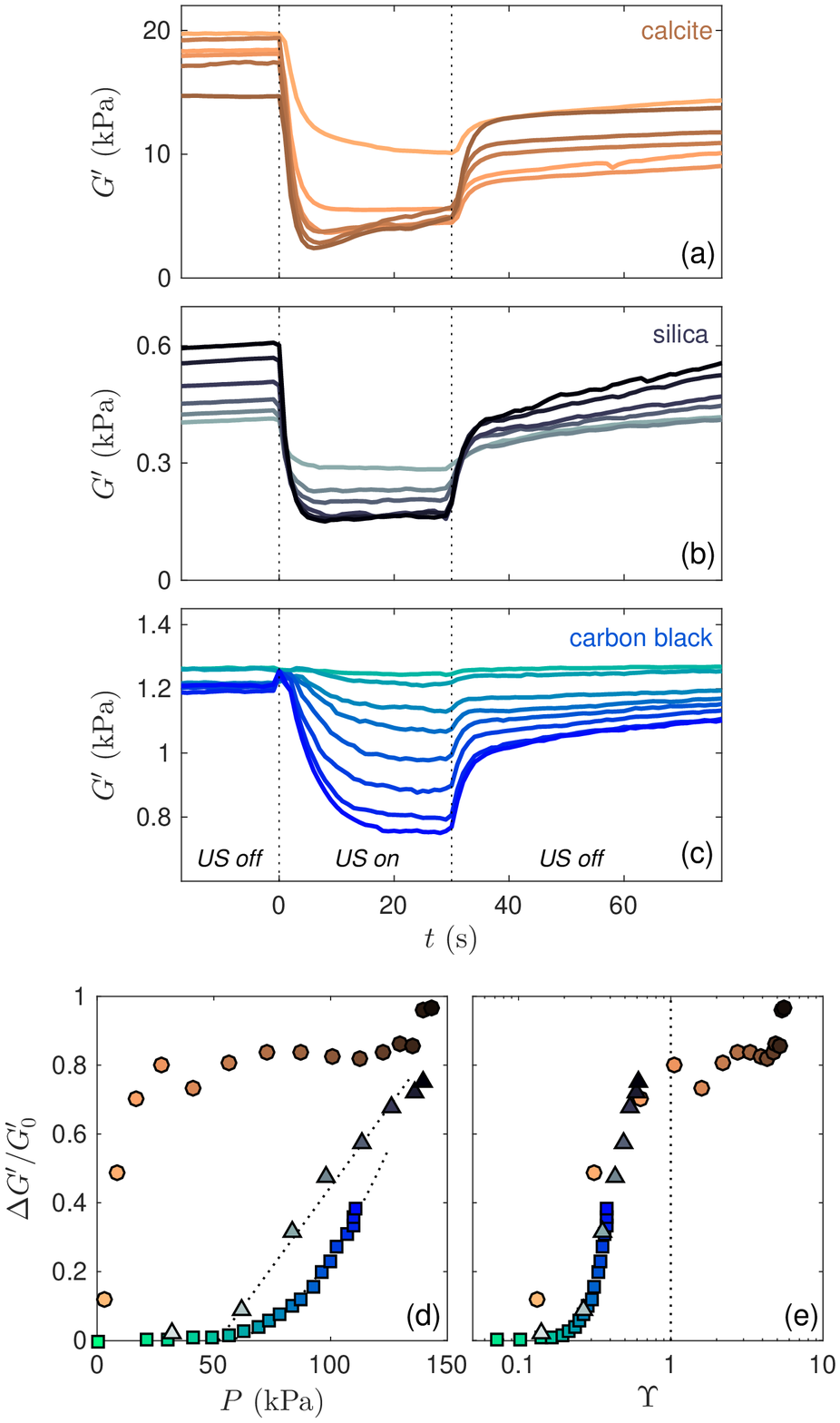}
         \caption{{\bf Ultrasound-induced softening of colloidal gels.} Elastic modulus $G'$ as a function of time $t$ for (a)~a 10~\% vol. calcite gel, (b)~a 5~\% vol. silica gel and (c)~a 3~\% vol. carbon black gel. Ultrasonic vibrations with frequency 45~kHz are turned on at time $t=0$ and switched off at $t=30$~s. Each curve corresponds to a given acoustic intensity, with darker colors corresponding to larger intensities. (d)~Relative amplitude of the softening effect $\Delta G'/G'_0$ as a function of the acoustic pressure $P$ for the calcite, silica and carbon black gels (from top to bottom). $\Delta G'=G'_0-\min(G')$ is the drop in elastic modulus induced by ultrasonic vibrations, with $G'_0=G(t=0)$ the initial value of the elastic modulus. The dotted lines emphasize the threshold at $P_\text{c}\simeq 50$~kPa for the silica and carbon black gels respectively through linear and quadratic behaviours in $P-P_\text{c}$. (e)~Relative softening amplitude $\Delta G'/G'_0$ as in (d) replotted as a function of the dimensionless group $\Upsilon=\gamma^\text{s}_\text{US}/\gamma_\text{NL}$ that compares the shear strain induced by ultrasound to the onset of nonlinear response at 1~Hz. The dotted line shows $\Upsilon=1$.}
     \label{fig:modulus}
 \end{figure}

The displacement of the transducer surface remains always much smaller than the gap size. At the highest achievable acoustic intensity, its maximum amplitude $a$ reaches $0.5~\mu$m for the 45-kHz transducer (as calibrated with a laser vibrometer, Polytec OVF-505) and up to $0.03~\mu$m for the 500-kHz transducer (as derived from the manufacturer calibration). Assuming spatially homogeneous oscillations, vibrations with amplitude $a$ generate an oscillatory extensional strain (along the direction of the vibrations i.e. along the vertical direction) of the order of $\gamma^\text{e}_\text{US}\sim a/h$ and a shear strain (along the horizontal direction) of the order of $\gamma^\text{s}_\text{US}\sim Ra/2h^2$, with $R$ the radius of the parallel-plate geometry and $h$ the gap width. Since $h\ll R$, $\gamma^\text{e}_\text{US}\ll\gamma^\text{s}_\text{US}$ and the maximum strain amplitude is $\gamma^\text{s}_\text{US}\simeq 0.4$~\% at 45~kHz and 0.04~\% at 500~kHz. The former order of magnitude is comparable to the onset of nonlinear viscoelasticity at low frequency ($\gamma_\text{NL}=0.1$--1\%) and always below the yield strain of the gels ($\gamma_\text{y}=1$--10\%) as obtained from the standard oscillatory shear measurements at 1~Hz displayed in Fig.~\ref{fig:strainsweep} in Appendix~\ref{app:gels}. Moreover, the amplitude of the effective strain rate associated with ultrasonic vibrations, $\dot\gamma^\text{s}_\text{US}=\omega\gamma^\text{s}_\text{US}$, reaches about 200~s$^{-1}$ for both ultrasonic frequencies. We emphasize that the high-frequency nonlinear rheology of our gels is unknown so that these estimates remain difficult to interpret. Still, based on these significant values for strain and strain rates, one may expect that ultrasonic vibrations should interact nonlinearly with the samples and have a strong impact on their mechanical properties.

Rather than in terms of displacement, the acoustic amplitude is usually reported in terms of the velocity $v$ or pressure $P=\rho c v=\rho c \omega a$, where $\rho$ is the gel density, $c$ is the speed of sound in the gel and $\omega=2\pi f$ is the angular frequency. In our experiments, $P$ ranges up to 150~kPa for both transducers. This corresponds to a maximum acoustic power of about $\mathcal{P}=P^2/\rho c\simeq 1.5$~W\,cm$^{-2}$.

\paragraph*{Results.}
We start by simply monitoring the viscoelastic moduli $G'$ and $G''$ under ultrasonic vibrations following the protocol detailed in Appendix~\ref{app:protocol_rheo}. Our first important result is that the elastic modulus $G'$ of a colloidal gel may dramatically drop under ultrasonic vibrations. As shown in Fig.~\ref{fig:modulus}, this softening effect increases with acoustic pressure $P$, up to drops in gel elasticity by a factor of about five at the highest pressures, indicative of ultrasound-induced collapse (see also Fig.~\ref{fig:modulus_normalized} in Appendix~\ref{app:softening}). The softening depends on the type of gel both qualitatively and quantitatively. Although the calcite gel in Fig.~\ref{fig:modulus}(a) has the largest elastic modulus at rest, it is the most fragile system and therefore turns out to be the most sensitive to ultrasonic vibrations by showing a rapid saturation of the relative softening amplitude with the acoustic pressure [Fig.~\ref{fig:modulus}(d)]. By contrast, the silica gel in Fig.~\ref{fig:modulus}(b) and the carbon black gel in Fig.~\ref{fig:modulus}(c), which respectively show strong and weak physical aging at rest and yield strains larger than the calcite gel (see Fig.~\ref{fig:strainsweep} in Appendix~\ref{app:gels}), require that ultrasonic vibrations reach some threshold amplitude in order to induce softening. 

In order to rationalize the above observations, we introduce the dimensionless group $\Upsilon=\gamma^\text{s}_\text{US}/\gamma_\text{NL}$ that compares the amplitude of the shear strain induced by ultrasound to the strain amplitude at onset of nonlinear response at 1~Hz inferred from Fig.~\ref{fig:strainsweep} in Appendix~\ref{app:gels}. As shown in Fig.~\ref{fig:modulus}(e), the softening amplitudes for all three gels appear to fall onto a single curve when the acoustic excitation is rescaled in terms of $\Upsilon$. The collapse is especially good for the carbon black and silica gels and points to a critical value of $\Upsilon_\text{c}\simeq 0.1$ below which ultrasonic vibrations have no measurable effect on the elastic modulus. Although more data at higher $\Upsilon$ values would be needed for both these gels, the calcite rescaled data indicate that the softening effect saturates for $\Upsilon\gtrsim 1$, which highlights the relevance of this dimensionless number. Indeed, we tried to rescale the same data using other dimensionless groups based on the gel yield stress $\sigma_\text{y}$, including $P/\sigma_\text{y}$ and $\eta_\text{f}P/\rho c d \sigma_\text{y}$, with $\eta_\text{f}$ the viscosity of the suspending fluid and $d$ the diameter of the colloidal particles (see also discussion in Section~\ref{sec:conclu}) but all these definitions led to scattered data sets for the three gels under study and/or to irrelevant values of the dimensionless parameter, either much larger or much smaller than 1 over the whole range of acoustic amplitudes.

Remarkably, after ultrasound is turned off, all gels recover most of their elasticity over time [see $t>30$~s in Figs.~\ref{fig:modulus}(a--c)]. This suggests that the effect of ultrasound on the colloid network is essentially reversible. For the silica and carbon black gels, some softening persists for large acoustic pressures while the calcite gel shows the reverse trend with full recovery for the largest pressures. Such a dependence with the gel microstructure is not surprising since a complex interplay between the ultrasonic vibration and the colloid network is expected at the microscale. This issue is investigated in more detail below in the case of the carbon black gel.

Since one may also invoke surface effects, such as partial detachment of the gel at the walls, to explain the softening reported in Fig.~\ref{fig:modulus}, it is crucial to demonstrate that ultrasonic vibrations actually induce \emph{bulk} modifications of the gel microstructure. This is what we proceed to show in the next section where time-resolved structural measurements are performed through ultra small-angle x-ray scattering (USAXS) as ultrasonic vibrations are being applied to a 4-mm slab of carbon black gel. 

\section{Gel structure under high-power ultrasonic vibrations}
\label{sec:saxs}

\paragraph*{Experimental setup.}
Our setup for time-resolved USAXS measurements under ultrasound is described in full details in Ref.~\cite{Jin:2014a}. Briefly, a titanium vibrating blade of width 2~mm connected to a piezoelectric transducer working at 20~kHz (SODEVA TDS, France) is immersed in a channel of width 4~mm, depth 8~mm and length 100~mm. Originally designed for cross-flow ultrafiltration, this setup is used here in the absence of any flow except that used for filling and preshearing the sample (see protocol in Appendix~\ref{app:protocol_saxs}). USAXS measurements are carried out with the TRUSAXS instrument at the ID02 High Brilliance beamline of the European Synchrotron Radiation Facility (ESRF, Grenoble, France) \cite{Narayanan:2018}. The incident x-ray beam of wavelength 0.1~nm is collimated to a vertical size of 80~$\mu$m and a horizontal size of 150~$\mu$m. A sample-to-detector distance of 31~m is used and provides access to scattering magnitudes $q$ of the scattering wave vector from 0.0008 to 0.155~nm$^{-1}$.

USAXS measurements are performed on the carbon black gel at a distance of about 1~mm from the vibrating blade through a small window of thickness 0.3~mm machined in the polycarbonate channel wall. The background scattering from the channel filled with the mineral oil used in the carbon black gel is systematically subtracted to the two-dimensional scattering patterns. The resulting scattering intensity is isotropic so that only radially averaged spectra are presented. Similar measurements on calcite and silica gels were not possible due to multiple scattering by the 4-mm thick samples.

 \begin{figure}
 	\centering
 	\includegraphics[width=\columnwidth]{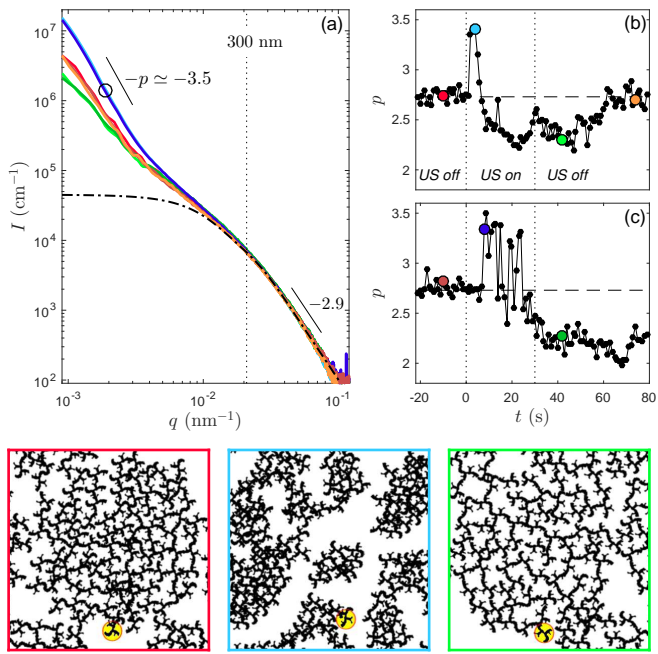}
     \caption{{\bf Structural measurements under ultrasonic vibrations.} (a) Time-resolved USAXS intensity spectra $I(q)$ recorded in a 3~\% vol. carbon black gel under ultrasonic vibrations with frequency 20~kHz for two different acoustic pressures $P=64$~kPa (light colors) and $P=240$~kPa (dark colors). Vibrations are turned on at $t=0$ and switched off at $t=30$~s. Three families of spectra can be distinguished as a function of time before (red), during (blue) and just after application of ultrasound (green). The times at which the spectra are shown correspond to the various colored symbols in (b,c). The black dash-dotted curve is the form factor used for the data analysis presented in Appendix~\ref{app:saxs}. The dotted line indicates the magnitude $q$ of the scattering wave vector that corresponds to 300~nm, the effective diameter of the carbon black particles. (b)~Exponent $p$ of the best power-law fit $I(q)\sim q^{-p}$ at low $q$ as a function of time for $P=64$~kPa. The structure fully recovers after ultrasonic vibrations are turned off [see orange dot and corresponding spectrum in (a)]. (c)~Same as (b) for $P=240$~kPa. The structure remains affected by ultrasound over at least one minute. Bottom panels: qualitative views of the gel microstructure before (left), during (middle) and after application of ultrasound (right) as inferred from USAXS data. The yellow circle highlights a carbon black particle of effective diameter 300~nm (scale bar).}
     \label{fig:saxs}
 \end{figure}

\paragraph*{Results.}
As displayed in Fig.~\ref{fig:saxs}(a), the scattering intensity $I(q)$ is unaffected by ultrasonic vibrations for scattering wave vector magnitudes $q\gtrsim 10^{-2}$~nm$^{-1}$, i.e., the structure of the carbon black gel does not change for length scales smaller than about 500~nm (see also Figs.~\ref{fig:saxs_fits_13pc} and \ref{fig:saxs_fits_100pc} in Appendix~\ref{app:saxs} for a more detailed analysis of the USAXS data). However, for $q\lesssim 10^{-2}$~nm$^{-1}$, the scattering intensity $I$ strongly varies upon application of ultrasound. To quantify this effect, we measure the power-law exponent $p$ of the scattering intensity, $I(q)\sim q^{-p}$ at low $q$ values. Starting from $p\simeq 2.7$ at rest, the exponent reaches values as large as 3.5 under medium-amplitude ultrasonic vibrations ($P=64$~kPa), before undershooting to $p\simeq 2.3$ and finally slowly relaxing to its initial value once vibrations are turned off [Fig.~\ref{fig:saxs}(b)]. Under a much higher acoustic pressure ($P=240$~kPa), series of bursts with peak values $p\simeq 3.5$ are observed and the intensity spectrum does not fully recover its initial shape at rest [Fig.~\ref{fig:saxs}(c)].

The above results not only allow us to unambiguously identify the effects of ultrasonic vibrations as bulk effects, they also lead us to associate them with a qualitative microscopic scenario. Indeed, exponents $p$ smaller than 3 indicate that the bulk structure of the material contributes the most to the scattering and that the scattered intensity corresponds to a {\it mass} fractal with fractal dimension $d_\text{f}=p$. On the other hand, exponents ranging between 3 and 4 indicate that interfaces dominate the scattering and such exponents are interpreted in terms of a {\it surface} fractal with fractal dimension $d_\text{f}=6-p$, with $p=4$ as the limiting case of a smooth surface \cite{Schmidt:1991}. Here, ultrasonic vibrations are observed to induce a transition from $p\simeq 2.7<3$ to $p\simeq 3.5>3$ for wave numbers which magnitudes range from the lowest achievable values up to $q\simeq 5\,10^{-3}$~nm$^{-1}$, i.e. over half a decade at low $q$. This implies that, for length scales $2\pi/q\gtrsim 1~\mu$m, scattering is dominated by a mass fractal, bulk structure at rest, which turns into a structure with fractal interfaces under ultrasound. Therefore, as pictured in the bottom panels of Fig.~\ref{fig:saxs}, the most probable scenario is that the gel network gets fractured by ultrasonic vibrations over length scales larger than typically 1~$\mu$m. The resulting cracks are filled with oil and the surface roughness of these channels accounts for the scattering with $p\simeq 3.5$. This fragmentation results in a drop of the gel elasticity. The system then coalesces back into a mass-fractal structure looser than the initial one ($p\simeq 2.3$). At high acoustic pressures, the large temporal fluctuations and the oscillations from surface to mass fractal hint at spatial heterogeneity and possible large-scale flow. After vibrations are turned off, the degree of structural recovery depends on the acoustic pressure consistently with the behaviour of the elastic modulus reported in Fig.~\ref{fig:modulus}(c).

\section{Gel flow and yield stress under high-power ultrasonic vibrations}
\label{sec:flow}

In order to deepen our exploration of the interaction between colloidal gels and ultrasonic vibrations, we now return to the rheo-ultrasonic setup used in Fig.~\ref{fig:modulus} and address the effect of vibrations on yielding, both through flow curve measurements (Fig.~\ref{fig:gdot_down}) and through creep experiments (Fig.~\ref{fig:fluidization}) following the protocols in Appendix~\ref{app:protocol_rheo}. 

\paragraph{Flow curves under ultrasonic vibrations.}
We focus on flow curves obtained on the calcite gel by measuring the shear stress $\sigma$ while sweeping down the shear rate $\dot\gamma$ for different acoustic pressures $P$. As seen in Fig.~\ref{fig:gdot_down}(a), all flow curves show the same general trend. At large $\dot\gamma$, the gel flows and the stress increases as $\sigma \sim \dot\gamma^n$ with $0<n<1$ the shear-thinning index. We observe a sharp increase of $n$ with $P$ [Fig.~\ref{fig:gdot_down}(c)] indicating that the calcite gel becomes less and less shear thinning as the intensity of ultrasonic vibrations is increased. Since the whole gel is fluidized at large shear rates and potential surface effects become negligible, this dependence of $n$ on $P$ further confirms the bulk nature of the effect of vibrations. At low $\dot\gamma$, the stress plateaus to a value that corresponds to the dynamic yield stress $\sigma_\text{y}$ \cite{Bonn:2017}. In the calcite gel, $\sigma_\text{y}$ is observed to decrease linearly by almost one order of magnitude over the explored range of acoustic pressures [Fig.~\ref{fig:gdot_down}(b)] indicating that increasing the acoustic intensity makes it easier to break and fluidize the gel. The very same trends are obtained from static yield stress measurements where the shear stress is swept up from the rest state (see Fig.~\ref{fig:stress_up} in Appendix~\ref{app:softening}). We conclude that, as one increases the acoustic energy injected into the material, ultrasonic vibrations can be used to turn a colloidal gel with a strong yield stress behaviour ($\sigma_\text{y}\simeq G_0$) into a more and more Newtonian fluid ($\sigma_\text{y}\ll G_0$ and $n\rightarrow 1$).

\begin{figure}
 	\centering
 	\includegraphics[width=\columnwidth]{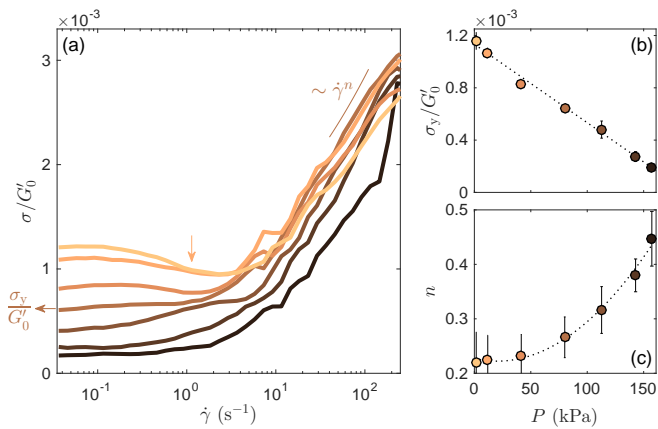}
     \caption{{\bf Flow properties under ultrasonic vibrations.} (a)~Flow curves, shear stress $\sigma$ vs shear rate $\dot\gamma$, measured in a 10~\% vol. calcite gel for different acoustic intensities [increasing from top to bottom, see color code in (b,c)]. The frequency of ultrasonic vibrations is 45~kHz. The shear stress is normalized by the elastic modulus $G'_0$ measured at rest. (b)~Normalized yield stress $\sigma_\text{y}/G'_0$ and (c)~shear-thinning exponent $n$ as a function of the acoustic pressure $P$. Dotted lines in (b) and (c) respectively show the best linear and parabolic fits of the data. The vertical arrow in (a) points to the decreasing region of the flow curve at low $P$.}
     \label{fig:gdot_down}
 \end{figure}

Furthermore, we note that in the absence of ultrasonic vibrations, the flow curve displays a decreasing region for $0.5\lesssim\dot\gamma\lesssim 5$~s$^{-1}$. Such an unstable branch is generically observed in attractive colloidal systems and interpreted as the signature of localized or shear-banded flows \cite{Bonn:2017,Divoux:2013}. Interestingly, under ultrasonic vibrations, this decreasing portion progressively gives way to a monotonically increasing branch as the amplitude of the vibrations is increased. Very similar features have been reported in vibrated frictional grains \cite{Dijksman:2011} as well as vibrated granular suspensions \cite{Hanotin:2012,Hanotin:2015}. This peculiar behavior was linked to the presence of a dynamical critical point at a finite flow rate due to the competition between flow-induced fluctuations and external vibration \cite{Wortel:2016}. Very recently, the previous approach developed for frictional grains was generalized to soft glassy materials through a mesoscale elastoplastic model accounting for a shear-banding instability and showing again a transition from flow curves with a minimum to monotonically increasing flow curves as vibration is increased \cite{LeGoff:2019}. The results of Fig.~\ref{fig:gdot_down} may constitute the first experimental evidence that such a scenario applies to colloidal gels.

 \begin{figure}
 	\centering
 	\includegraphics[width=\columnwidth]{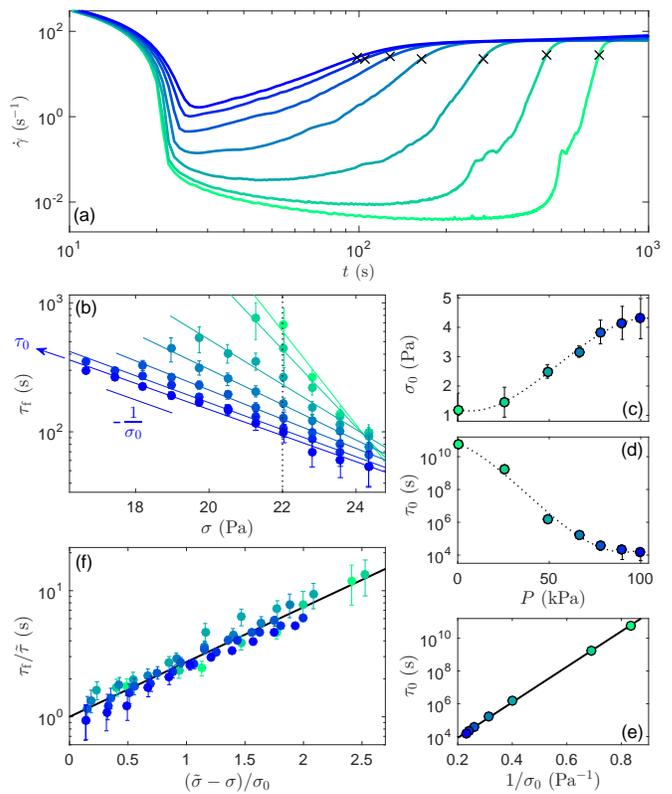}
     \caption{{\bf Ultrasound-assisted fluidization.} (a)~Shear rate responses $\dot\gamma(t)$ measured in a 3~\% vol. carbon black gel for different acoustic intensities [increasing from right to left, see color code in (c,d)] after a constant shear stress $\sigma=22$~Pa is applied following preshear. The frequency of ultrasonic vibrations is 45~kHz. Following a creep period, the gel starts to flow within a well-defined fluidization time $\tau_\text{f}$ (see black $\times$) that sharply decreases with the acoustic intensity. (b)~Fluidization time $\tau_\text{f}$ as a function of the applied stress $\sigma$ under different acoustic intensities (increasing from top to bottom). Colored solid lines are the best exponential fits, $\tau_\text{f}=\tau_0\exp(-\sigma/\sigma_0)$, for a given acoustic intensity. The vertical dotted line shows the shear stress $\sigma=22$~Pa used in (a). (c)~Characteristic stress $\sigma_0$ and (d)~time $\tau_0$ extracted from exponential fits of $\tau_\text{f}(\sigma)$ as a function of the acoustic pressure $P$. Dotted lines are polynomial fits to guide to the eye. (e)~$\tau_0$ as a function of $1/\sigma_0$ in semilogarithmic scales. The solid line is the best exponential fit $\tau_0=\tilde{\tau}\exp(\tilde{\sigma}/\sigma_0)$ with $\tilde{\tau}=57.1$~s and $\tilde{\sigma}=24.9$~Pa. (f)~Master curve for fluidization times $\tau_\text{f}/\tilde{\tau}$ as a function of $(\tilde\sigma-\sigma)/\sigma_0$. The black line is the exponential function $y=\exp(x)$.}
     \label{fig:fluidization}
 \end{figure}

\paragraph{Ultrasound-induced acceleration of fluidization under stress.}
In the experiments of Fig.~\ref{fig:gdot_down}, the flow curves could be reasonably interpreted as steady states thanks to the weak time-dependence of the calcite gel. However, when made to flow, colloidal gels generically show strong time-dependence, through e.g. thixotropy or rheopexy, due to the competition between physical aging and shear-induced rejuvenation \cite{Mewis:2009,Radhakrishnan:2017}. Carbon black gels enter such a category. More specifically,  carbon black gels are known to be prototypical of the so-called ``delayed yielding'' phenomenon: during a creep experiment, when submitted to a constant stress $\sigma$ higher than, yet close to, the yield stress, the presheared gel first consolidates through a long-lasting creep regime before getting fluidized after a time $\tau_f$ that follows an exponential decrease with the applied stress, $\tau_\text{f}=\tau_0\exp(-\sigma/\sigma_0)$, characteristic of activated processes \cite{Gibaud:2010,Sprakel:2011,Ovarlez:2013,Grenard:2014}. Figure~\ref{fig:fluidization} summarizes the effect of ultrasonic vibrations on such delayed fluidization. First, for a given applied stress, vibrations are seen to accelerate fluidization by a factor of about ten at the highest achievable acoustic pressure [Fig.~\ref{fig:fluidization}(a)]. Varying the applied stress for a fixed acoustic pressure $P$ [Fig.~\ref{fig:fluidization}(b)], we recover the classical exponential behaviour for $\tau_\text{f}$ vs $\sigma$, yet with parameters $\sigma_0$ and $\tau_0$ that strongly depend on $P$ [Fig.~\ref{fig:fluidization}(c,d)]. Mean-field approaches of delayed yielding predict that (i)~$\sigma_0\sim\rho_0 k_\text{B}T/\delta$, where $\rho_0$ is the initial area density of strands, $k_\text{B}$ the Boltzmann constant, $T$ the temperature and $\delta$ the range of the interaction potential and (ii)~$\tau_0\sim\exp(E_\text{A}/k_\text{B}T)$ where $E_\text{A}$ is the depth of the potential well \cite{Lindstrom:2012}. Since ultrasonic vibrations are unlikely to affect $\rho_0$, $\delta$ or $E_\text{A}$, a way to interpret our results is to invoke an \emph{effective} temperature $T_\text{eff}\propto\sigma_0$ associated with the vibrational ultrasonic energy, as commonly done for agitated granular materials \cite{Jaeger:1996,dAnna:2003,Puglisi:2005}. As discussed in detail in Appendix~\ref{app:Teff}, the remarkable Arrhenius-like scaling of $\tau_0$ with $\sigma_0$ demonstrated in Fig.~\ref{fig:fluidization}(e) further allows us to collapse all the fluidization time data onto a single exponential master curve in Fig.~\ref{fig:fluidization}(f) and to extract an estimate of the interaction energy $E_\text{A}\simeq 20.8\,k_\text{B}T$ that is consistent with previous work \cite{Trappe:2001,Prasad:2003,Trappe:2007}. This fully confirms that the characteristic stress $\sigma_0$ measures the effective temperature of the system, which increases by a factor of about 4 over the range of acoustic intensities explored here. Therefore, injecting acoustic energy into the system amounts to increasing its temperature and thus facilitates the fluidization of gels that display delayed yielding.

\section{Discussion and conclusion}
\label{sec:conclu}

To summarize, the three colloidal gels investigated here can be called ``rheo-acoustic'' gels as their mechanical and flow properties display great sensitivity to ultrasonic vibrations. We have shown that ultrasonic vibrations transiently soften these fragile materials and facilitate their flow.  This concept is all the more promising that the gels we chose to prove our point, calcite, silica or carbon black gels, span a large range of yielding behaviors usually observed in soft solids. We have quantified these effects thanks to setups that couple a built-in ultrasonic transducer to rheology and USAXS measurements and using protocols for gels at rest, under shear flow and submitted to a constant stress. These effects are robust and persist through changes in the ultrasound frequency (Figs.~\ref{fig:modulus_frequency} and \ref{fig:fluidization_frequency} in Appendix~\ref{app:frequency}) and in the cell geometry (Fig.~\ref{fig:modulus_geometry} in Appendix~\ref{app:geometry}). Therefore, they should swiftly become key to applications in flow enhancement, unclogging, extrusion or 3D-printing based on acoustically-tunable colloid-based materials.

The present results also entail a number of new fundamental questions that should further pave the way for future theoretical and numerical work. At a mean-field level, our fluidization results strongly suggest that some of the mechanical effects observed here can be rationalized through an effective temperature. This corroborates the general idea that external stress or perturbations can be seen as a ``mechanical noise'' that adds up to thermal noise. Experimentally, one could deepen the analogy with real temperature by performing fluidization experiments as in Fig.~\ref{fig:fluidization} over a wide range of temperatures in the absence of ultrasonic excitation and by checking whether the acceleration of the dynamics follows the same trend as under ultrasound. As for theory, the mesoscopic approach proposed in Ref.~\cite{LeGoff:2019} is based on the assumption that vibrations induce a ``fluidizing noise'' that leads to additional activated plastic events of the same nature as those induced by shear \cite{Picard:2005} but with a ``vibration rate'' $k_\text{vib}$ that quantifies the influence of vibrations. Following this kind of approach to predict viscoelastic properties of soft glassy materials under vibrations or to study their effect on yielding dynamics looks very promising for future comparison with the present experiments. Moreover, extending classical theoretical approaches such as the Soft Glassy Rheology framework \cite{Sollich:1997,Fielding:2000}, that relies on a ``noise temperature,'' or Shear-Transformation Zone models \cite{Falk:1998,Falk:2011}, which involve an effective ``disorder temperature,'' to include the effect of mechanical vibrations would be most insightful.

Besides the modelling of the global signature of ultrasonic vibrations on mechanical properties, our experiments raise important physical questions about the interaction mechanism between colloidal gels and ultrasonic vibrations at a microscopic level. First, a few ultrasound--gel interaction mechanisms can be disregarded. Indeed, at the rather low acoustic powers involved in our study (${\mathcal{P}}\lesssim 2$~W\,cm$^{-2}$), cavitation bubbles are very unlikely to occur \cite{Jin:2014a}, especially in the oil-based carbon black gel, and any influence of sonochemistry can be excluded \cite{Chandrapala:2012}. The acoustic wavelength (3--60~mm) being always several orders of magnitude larger than the microstructure characteristic sizes (20~nm--1~$\mu$m), radiation pressure through scattering of ultrasound by colloidal clusters may also be neglected \cite{Hamilton:1998}. 

Second, the scaling in terms of the parameter $\Upsilon=\gamma^\text{s}_\text{US}/\gamma_\text{NL}$ reported in Fig.~\ref{fig:modulus}(e) suggests that at least the softening effect is controlled by the {\it strain} induced by ultrasonic vibrations. Indeed, it seems natural to think of ultrasonic vibrations as a source of oscillatory mechanical stress analogous to that of oscillatory shear although not through simple shear flow. Repetitively applying the extensional and shear strains $\gamma^\text{e}_\text{US}$ and $\gamma^\text{s}_\text{US}$ introduced in Sect.~\ref{sec:softening} at high frequency could induce fatigue in the gel and therefore its softening or even its fluidization. Such fatigue could explain that the effects of ultrasonic vibrations become noticeable as soon as $\Upsilon\gtrsim 0.1$, far below the onset of nonlinearity in the rheological response. Note, however, that ultrasonic vibrations impose oscillations along the vorticity direction at a high frequency ($>20$~kHz) whereas the rheometer imposes a constant or slowly varying shear (typically $<20$~Hz). Such a wide gap in frequencies makes any direct analogy with the low-frequency response hypothetical and it is not clear how standard rheological observables such as the elastic modulus or the yield stress extrapolate to the ultrasonic frequency domain.

Moreover, while the above argument based on strain-induced fatigue seems relevant for explaining the softening effect at rest, it is unlikely to also hold under flow. This raises the question of the existence of a single dimensionless number that would govern {\it all} the effects of ultrasonic vibrations on colloidal gels. Such a dimensionless group should compare vibration-induced forces to internal forces within the gel. Still, depending on whether one considers gels at rest or under shear, one may propose to base vibration-induced forces on elastic forces or on viscous friction, either at the single particle level or at the gel network level. As for internal forces, these obviously result from interparticle attraction but may depend in a complex manner on the microscopic details of such attraction, on the fractal geometry of the network, on internal stresses stored during preshear and on thermal energy. 

At this stage, it is interesting to compare the present study on colloidal gels to previous findings on dense suspensions. Recently, shear thickening in concentrated Brownian and non-Brownian suspensions could be tuned both through the superposition of an oscillatory flow with frequency up to 500~Hz to steady shear \cite{Lin:2016b} and through the application of acoustic perturbations \cite{Sehgal:2019}, a process that was shown to be rate-controlled. Similarly, friction and liquefaction in both dry and wet granular materials under low-frequency mechanical vibrations appear to be driven by the velocity amplitude $v=\omega a$ \cite{Hanotin:2012,Lastakowski:2015}. These effects have been interpreted in terms of consolidated {\it vs} mobile states depending on whether or not grains are in contact through force chains \cite{Hanotin:2015}, and further rationalized in terms of the lubrication Peclet number, $Pe_\text{lub}=\eta_\text{f}\omega a/\mu P_\text{g}d$, where $\eta_\text{f}$ is the viscosity of the suspending fluid, $\mu$ the effective friction coefficient, $P_\text{g}$ the granular pressure and $d$ the bead diameter \cite{Hanotin:2012,Hanotin:2015}. This dimensionless number compares the viscous forces induced by vibrations to the internal friction forces in the granular system. An equivalent number for colloidal gels is $\Pi=\eta_f\omega a / d\sigma_\text{y}=\eta_\text{f}P/\rho c d \sigma_\text{y}$. Measurements on the calcite gel at two different ultrasonic frequencies (Fig.~\ref{fig:modulus_frequency} in Appendix~\ref{app:frequency}) suggest that $v=\omega a=P/\rho c$ could be the relevant control parameter but the softening data of Fig.~\ref{fig:modulus}(d) do not collapse as a function of $\Pi$. Therefore, future experimental studies should systematically explore the effects of the vibration frequency on gels with different volume fractions and for the various protocols. These studies will tell whether a universal dimensionless group involving $v$ (or equivalently $P$) or $a$ (or equivalently $\gamma^\text{s}_\text{US}$) and based on the yield strain or yield stress can generally describe the effects of acoustic perturbations on colloidal gels.

Finally, a detailed microscopic scenario remains to be uncovered. USAXS measurements unambiguously reveal that the gel network gets fractured by ultrasonic vibrations. Yet, structural measurements at larger length scales, e.g. through light scattering, are needed to get a full picture of the gel microstructure under ultrasonic vibrations. Moreover, in the absence of shear, we could not reach fully fluidized states (characterized by $G''>G'$) even though we pushed our transducers to the maximum achievable acoustic intensities. This suggests that, in spite of the presence of cracks and of possible large-scale rearrangements, some connectivity remains in the gel network so that solidlike behavior dominates. Future experimental work, if possible under even higher acoustic intensities, will focus on colloidal gels that can be imaged under a microscope in order to directly follow the breakup of the gel microstructure through high-speed imaging. At a mesoscopic level, particle imaging velocimetry or ultrasonic velocimetry will be helpful to investigate the possibility of vibration-induced wall slip and to confirm that shear-banding instabilities can be tuned by ultrasonic vibrations \cite{LeGoff:2019}. Deeper rheological investigations are also planned to explore the influence of previous shear history \cite{Ovarlez:2013,Helal:2016,Hipp:2019} on the ultrasound-induced effects evidenced here as well as hysteresis phenomena in flow curve measurements under ultrasound \cite{Divoux:2013,Radhakrishnan:2017}. We also believe that recent simulation methods developed for colloidal gels \cite{Colombo:2013,Varga:2015,Landrum:2016,Bouzid:2018}, once adapted to account for ultrasonic excitation, will certainly bring valuable insight into rheo-acoustic gels by investigating the local impact of ultrasonic vibrations on the gel microstructure and its relationship with the global mechanical response.

\section*{Acknowledgements}
The authors thank T.~Divoux and E.~Freyssingeas for insightful discussions on rheological and USAXS data, W.~Chevremont and T.~Narayanan for technical help with the USAXS--ultrasound combined setup, H.~Bodiguel and C.~de Loubens for sharing their beam time at ESRF with us, and C.~Barentin and T.~Liberto for providing us with calcite. This work was funded by the R\'egion Auvergne-Rh\^one-Alpes ``Pack Ambition Recherche'' Programme and by the European Research Council under the European Union's Seventh Framework Programme (grant agreement No. 258803). LRP is part of Labex TEC 21 (Investissements d'Avenir, grant agreement No. ANR-11-LABX-0030), PolyNat Carnot Institute (Investissements d'Avenir, grant agreement No. ANR-11-CARN-030-01) and of IDEX UGA program (ANR-15-IDEX-02). 

\appendix
\section{}

\subsection{Colloidal gels}
\label{app:gels}

\begin{figure}
 	\centering
 	\includegraphics[width=0.4\textwidth]{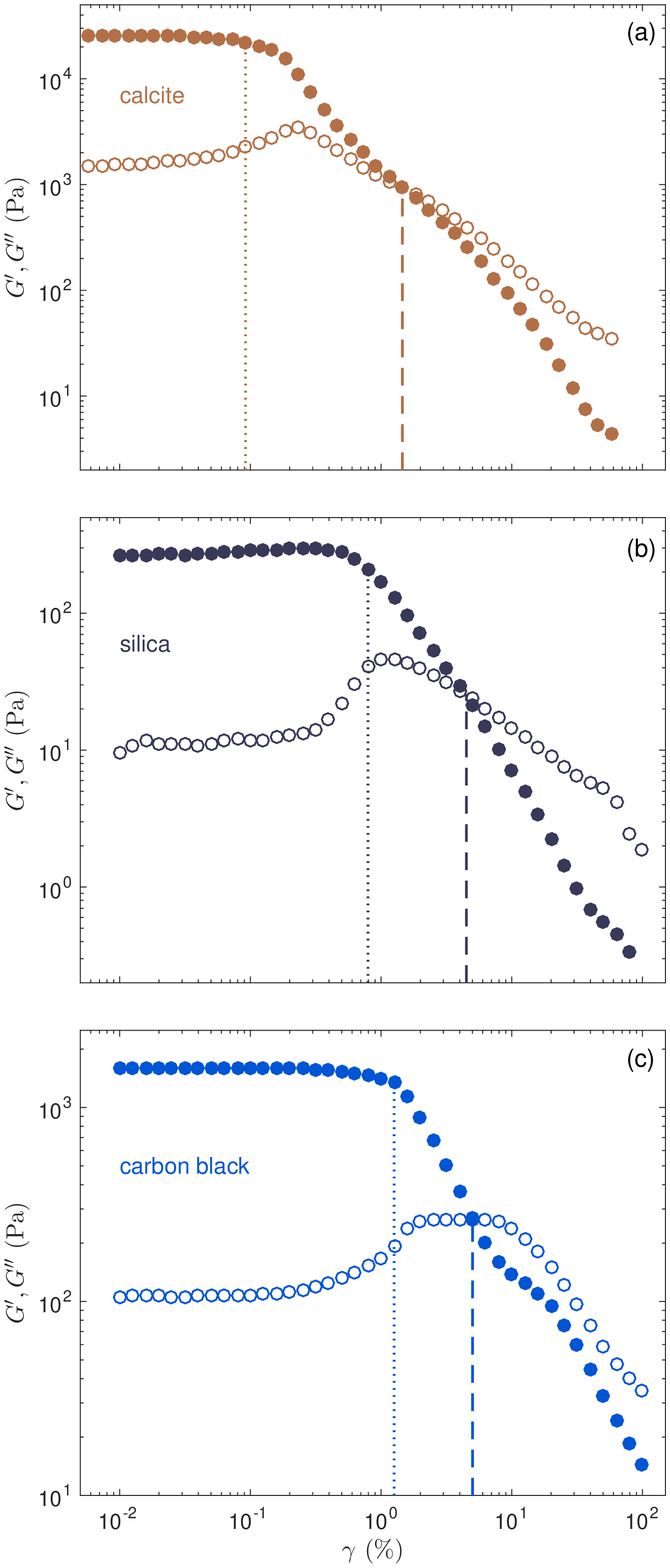}
     \caption{Elastic modulus $G'$ (filled symbols) and viscous modulus $G''$ (open symbols) as a function of strain amplitude $\gamma$ for (a)~a 10~\% vol. calcite gel, (b)~a 5~\% vol. silica gel and (c)~a 3~\% vol. carbon black gel. After the preshear and rest protocol described for each system in Appendix~\ref{app:gels}, the strain amplitude $\gamma$ is logarithmically swept up with 10 steps per decade and 10~s per point. The frequency of the oscillatory strain is 1~Hz. The dotted lines show the strain $\gamma_\text{NL}$ at which the elastic modulus falls below 85~\% of its value in the linear regime and which we take as the lower limit of the nonlinear regime. The dashed lines show the yield strain $\gamma_\text{y}$ defined as the point where $G'$ and $G''$ cross each other. Experiments conducted in a parallel-plate geometry of gap 1~mm. The upper plate is sandblasted in (a) and covered with sandpaper in (b,c).}
     \label{fig:strainsweep}
 \end{figure}

We focus on three different colloidal suspensions respectively made of calcite, silica and carbon black particles. All systems aggregate into space-spanning networks thanks to attractive interparticle forces. However, the resulting gels present different mechanical properties depending on the details of the interactions, as illustrated in Fig.~\ref{fig:strainsweep}.

Calcite is the most stable polymorph of calcium carbonate (CaCO$_3$). Calcite powder (Socal 31 from Solvay, average particle diameter of 60~nm, density of 2710~kg\,m$^{-3}$) is suspended in water at a volume fraction of 10~\%. The suspension is homogenized following the protocol described in Ref.~\cite{Liberto:2017} and spontaneously gels within a few minutes at pH~$\simeq 8.8$. The resulting network shows little aging over time yet great sensitivity to shear. A mild preshear is applied for 10~s at 10~s$^{-1}$ prior to any measurement. After a rest period of 60~s, the elastic modulus at rest reaches a steady value of about 20~kPa. 

Silica particles (SiO$_2$, Ludox TM~50 from Sigma, average particle diameter of 22~nm, density of 2350~kg\,m$^{-3}$) are made to aggregate at pH~$\simeq 9$ by mixing equal volumes of the 50~\% vol. stock Ludox solution and a concentrated NaCl solution to a final salt concentration of 2~M and final solid volume fraction of 5~\% as described in Ref.~\cite{Condre:2007}. Screening of the electrostatic repulsion leads to fast gel formation through attractive van der Waals forces and interparticle siloxane bonds \cite{Depasse:1997,Drabarek:2002}. Such a gel shows pronounced physical aging over time \cite{Kurokawa:2015}. Here, we minimize the effect of aging thanks to a strong preshear at 500~s$^{-1}$ applied for 120~s that partially erases the previous history. The gel is then left to rest for 300~s, after which the elastic modulus reaches about 500~Pa. As seen in Fig.~\ref{fig:modulus_normalized}(b), the systematic increase of the elastic modulus with successive experiments indicates that preshear does not allow for full rejuvenation of the sample. 

Carbon black particles (Cabot Vulcan XC72R, density of 1800~kg\,m$^{-3}$) are fractal aggregates of radius of gyration 100--200~nm and made of fused primary particles of typical  diameter 40~nm \cite{Richards:2017}. When dispersed in a mineral oil (density 838~kg\,m$^{-3}$, viscosity 25~mPa\,s, Sigma Aldrich), they aggregate via van der Waals attraction \cite{Hartley:1985} and form gels that show weak aging over time yet strong time-dependence under shear, including rheopexy \cite{Ovarlez:2013}, delayed yielding \cite{Gibaud:2010,Grenard:2014} and fatigue \cite{Gibaud:2016}. It was also shown that the preshear history can be used to tune the initial microstructure of carbon black gels \cite{Osuji:2008b,Ovarlez:2013,Helal:2016}. Here, we focus on a dispersion of carbon black particles at a volume fraction of 3~\% (weight fraction of 6~\% w/w). Because of the fractal nature of the particles, this corresponds to an effective volume fraction $\phi_\text{eff}\simeq 20$~\% \cite{Trappe:2007}. The preshear protocol is made of two successive steps of 200~s each, a first step at $-1000$~s$^{-1}$ in the ``reverse'' direction followed by one step at $+1000$~s$^{-1}$ in the ``positive'' direction that defines the direction of positive strains and stresses. After preshear, the dispersion recovers an elastic modulus of typically 1~kPa within a few seconds. 

The nonlinear viscoelasticity of the three colloidal gels under study is probed in the absence of ultrasonic vibrations by imposing an oscillatory shear with a fixed frequency of 1~Hz and an amplitude that increases from about 0.01~\% to 100~\%. As seen in Fig.~\ref{fig:strainsweep}, both the elastic modulus $G'$ and the viscous modulus $G''$ first remain independent of the oscillatory strain amplitude $\gamma$, which corresponds to linear response. We consider that the linear regime ends at a strain $\gamma_\text{NL}$ when the elastic modulus becomes smaller than 85~\% of its value in the linear regime (see dotted lines) and that the yield point is reached at the strain $\gamma_\text{y}$ when $G'=G''$ (see dashed lines). In the calcite gel, the linear regime extends only up to strain amplitudes of $\gamma_\text{NL}=0.09$~\% and the yield strain is $\gamma_\text{y}=1.5$~\% [Fig.~\ref{fig:strainsweep}(a)]. This behaviour was classified as that of a fragile gel in the ``strong link'' regime \cite{Shih:1990,Liberto:2017}. The linear regime of the silica gel extends up to $\gamma_\text{NL}=0.8$~\% and the gel yields at $\gamma_\text{y}=4.5$~\% [Fig.~\ref{fig:strainsweep}(b)]. Similarly, the response of the carbon black gel remains linear up to $\gamma_\text{NL}=1.3$~\% and the gel yields at $\gamma_\text{y}=5$~\% [Fig.~\ref{fig:strainsweep}(c)]. Therefore, in spite of a larger elastic modulus, the calcite gel is significantly more fragile than both the silica and the carbon black gels.

Finally, since the colloidal particles and the suspending fluids considered here are not density matched, sedimentation may affect the gel structure over long time scales. We estimate the gravitational Peclet number as defined in Ref.~\cite{Kim:2013}:
\begin{equation}
Pe_\text{g}=\frac{4\pi\vert\Delta\rho\vert g }{3k_\text{B}T}\,r^4\phi^{(d_\text{f}+1)/(d_\text{f}-3)}\,,
\end{equation}
where $\Delta\rho$ is the density difference between the particles and the suspending fluid, $g$ is the gravitational acceleration, $r$ is the particle radius and $\phi$ is the volume fraction. For the calcite gel, using a fractal dimension $d_\text{f}=2.23$ as determined in Ref.~\cite{Liberto:2017} leads to $Pe_\text{g}\simeq 0.2\lesssim 1$. In the case of the silica gel, assuming that aggregation is driven by a DLCA process for which $d_\text{f}\simeq 1.9$ is expected \cite{Condre:2007}, we estimate $Pe_\text{g}\simeq 5.10^{-4}\ll 1$. For the carbon black gel, giving an estimate of $Pe_\text{g}$ is more difficult since carbon black particles are themselves fractal aggregates and thus far from ideal spheres. Adapting the approach of Ref.~\cite{Kim:2013} to the case of fractal aggregates of radius $r$ and made of primary particles with radius $r_\text{p}$ with a fractal dimension $d_\text{p}$, one finds: 
\begin{equation}
Pe_\text{g}=\frac{4\pi\vert\Delta\rho\vert g }{3k_\text{B}T}\,r_\text{p}^{3-d_\text{p}} r^{d_\text{p}+1}\phi_\text{eff}^{(d_\text{f}+1)/(d_\text{f}-3)}\,,
\end{equation}
where $\phi_\text{eff}$ is the effective (hydrodynamic) volume fraction of the fractal aggregates. Taking $r_\text{p}=20$~nm and $d_\text{p}=2.7$ as reported in Ref.~\cite{Richards:2017} for Vulcan carbon black and $r=150$~nm and $d_\text{f}\simeq 2.6$ as inferred from USAXS measurements, one gets $Pe_\text{g}\simeq 5.10^{3}\gg 1$ for which gravitationally facilitated phase separation is expected. All our measurements are taken only 300~s after preshear (see protocols below in Appendix~\ref{app:protocol_rheo}) so that our data are most likely not affected by gravity whatever the gel. Yet, we note that $Pe_\text{g}$ spans many orders of magnitude covering both $Pe_\text{g}\ll 1$ and $Pe_\text{g}\gg 1$. This emphasizes the robustness of the effects such as the ultrasound-induced softening evidenced in our work while calling for a future detailed study of long-term gravitational effects under ultrasonic vibrations. Such a study could indeed provide a potential explanation for some of the discrepancies observed in Fig.~\ref{fig:modulus} between the various gels.

\subsection{Experimental protocols}

\subsubsection{Mechanical measurements under ultrasonic vibrations}
\label{app:protocol_rheo}

After the preshear protocol described above for each gel, the elastic modulus is monitored in our rheo-ultrasonic setup during 300~s through small-amplitude oscillatory strain (SAOS) of amplitude 0.06~\% and frequency 1~Hz. This allows us to define the ``initial'' elastic modulus $G'_0$ prior to application of vibrations.

In Fig.~\ref{fig:modulus}, ultrasonic vibrations are then applied for 30~s while measuring the elastic modulus with the same SAOS protocol as before (see also Fig.~\ref{fig:modulus_normalized}). Once vibrations are turned off, SAOS is continued for at least 170~s to monitor the recovery. For the flow curve measurements shown in Fig.~\ref{fig:gdot_down}, ultrasonic vibrations are applied while the shear rate is logarithmically swept down from 300 to 0.03~s$^{-1}$ within 80~s. For Fig.~\ref{fig:stress_up}, the shear stress is linearly swept up from 0 to 50~Pa in 60~s. Finally, for the creep experiments of Fig.~\ref{fig:fluidization}, both the target shear stress $\sigma$ and ultrasonic vibrations are applied just after preshear. Following Refs.~\cite{Sprakel:2011,Grenard:2014}, the fluidization time $\tau_\text{f}$ is defined as the last inflexion point of the shear rate response $\dot\gamma(t)$.

In all cases, the preshear protocol is repeated between successive tests at different acoustic intensities. We measured that after one minute under ultrasound, the temperature of the transducer surface (and thus of the sample) increases by less than 0.1$^\circ$C for $P<80$~kPa, by about 0.2 to 1$^\circ$C for $80<P<130$~kPa and by up to 4$^\circ$C for $P>130$~kPa. For fluidization experiments in the carbon black gel, which require to apply ultrasonic vibrations for much longer times, $T$ increases by 2 to 10$^\circ$C and the system is left to relax for one hour after sonication in order for the temperature to go back to its initial value. It is checked in Fig.~\ref{fig:temperature} in Appendix~\ref{app:temperature} that temperature changes cannot be invoked to explain the softening of colloidal gels observed under ultrasonic vibrations. Finally, we check in Fig.~\ref{fig:control} in Appendix~\ref{app:control} that rheological measurements are not perturbed by the application of ultrasonic vibrations.

\subsubsection{Microstructural measurements under ultrasonic vibrations}
\label{app:protocol_saxs}

The feed channel of our USAXS--ultrasound combined setup is filled with the gel under study thanks to a syringe pump (Harvard Apparatus PHD 4400). The gel is then presheared by pumping a volume of 10~mL at a flow rate of 20~mL\,min$^{-1}$ in both directions. Assuming a Poiseuille flow, this corresponds to a wall shear rate of about 8~s$^{-1}$ applied for 60~s. After a rest time of 180~s, ultrasonic vibrations are applied for 30~s with a pressure amplitude ranging from 20 to 240~kPa. USAXS spectra are recorded every second during 100~s starting 20~s prior to application of vibrations. Thus, these measurements provide insight into the microstructure under ultrasonic vibrations for acoustic powers similar to those of previous rheo-ultrasonic experiments.

\subsection{Ultrasound-induced softening: full data set}
\label{app:softening}
 
\begin{figure*}[!ht]
 	\centering
 	\includegraphics[width=0.8\textwidth]{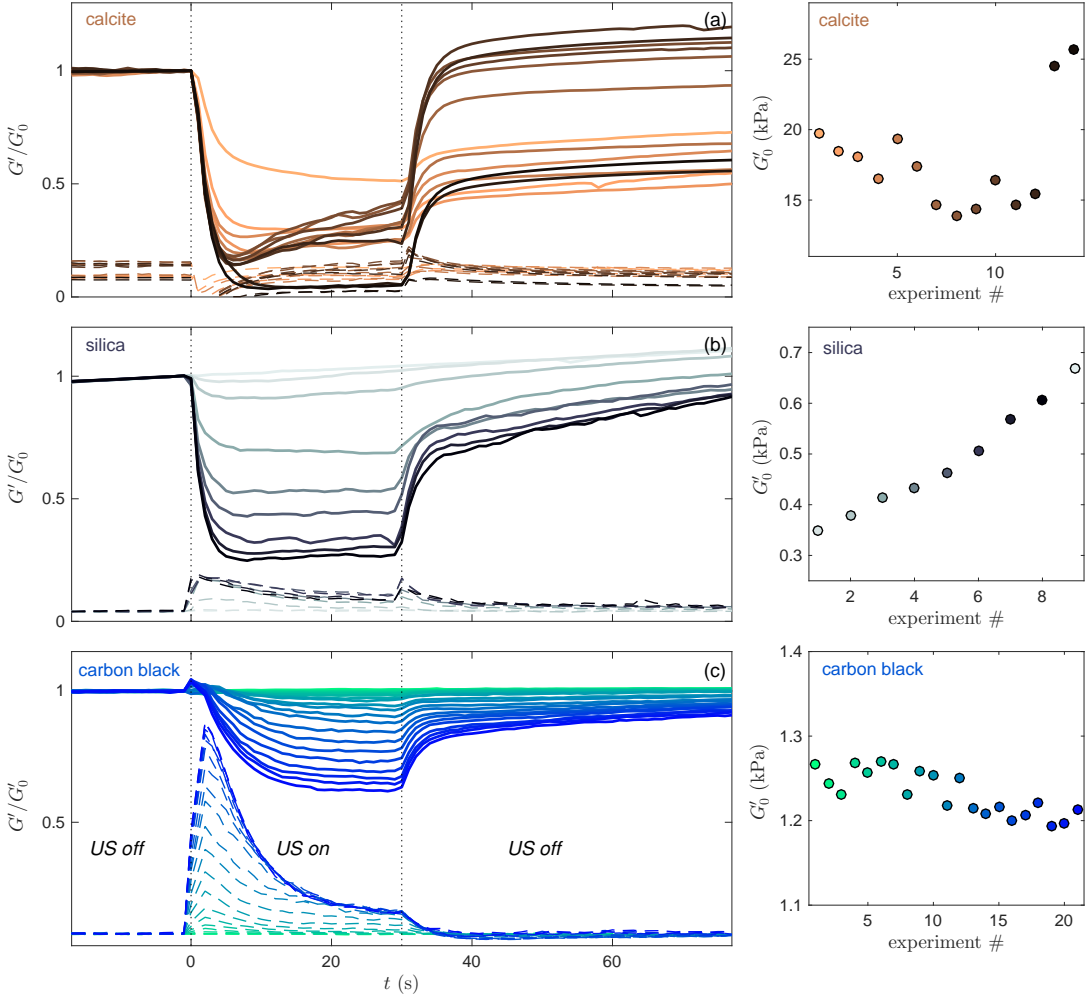}
     \caption{Elastic modulus (thick solid lines) and viscous modulus (thin dashed lines) normalized by the elastic modulus $G'_0$ measured just before ultrasonic vibrations are turned on and plotted as a function of time $t$ for (a)~a 10~\% vol. calcite gel, (b)~a 5~\% vol. silica gel and (c)~a 3~\% vol. carbon black gel. Ultrasonic vibrations with frequency 45~kHz are turned on at time $t=0$ and switched off at $t=30$~s. Each curve corresponds to a given acoustic intensity, with darker colors corresponding to larger intensities [see color code in Fig.~\ref{fig:modulus}(d)]. The right panels show the initial values $G'_0$ of the elastic modulus measured for the successive experiments performed under different acoustic intensities and used for normalization in the left panels.}
     \label{fig:modulus_normalized}
 \end{figure*}
 
Figure~\ref{fig:modulus_normalized} shows the viscoelastic moduli normalized by the elastic modulus $G'_0$ measured just before ultrasonic vibrations are turned on. The elastic (viscous resp.) modulus is plotted in solid lines (dashed lines resp.) for all available acoustic intensities and for the three colloidal gels under study in Fig.~1 in the main text. The right panels show that the initial modulus $G'_0$ displays a clear increasing trend with successive experiments only for the silica gel, i.e. for the system with the most pronounced aging effects. The last two measurements on the calcite gel [black curves in Fig.~\ref{fig:modulus_normalized}(a)] may have been affected by evaporation. Remarkably, the viscous modulus never exceeds the elastic modulus for all systems, even transiently. This means that although the gels display a spectacular softening, they remain solidlike and never reach a fully fluidized state under ultrasonic vibrations at the acoustic intensities explored here for a frequency of 45~kHz.

 \subsection{Analysis of USAXS measurements}
 \label{app:saxs}
 
 In order to go beyond the simple power-law description of the intensity spectra $I(q)$ proposed in the main text, we model the USAXS data obtained on the 3~\% vol. carbon black gel with the Beaucage model \cite{Beaucage:2012}. Originally designed for polymeric mass fractals \cite{Beaucage:1995,Beaucage:1996}, this model uses a ``unified'' function that interpolates between a Guinier regime at low $q$ and a power-law Porod scaling at large $q$. The crossover between the two regimes is controlled by a characteristic length scale $R_\text{g}$. In order to account for successive structural levels covering different ranges of scales, experimental data are often fitted to sums of such unified functions. Here, we describe our USAXS spectra by using two structural levels as contained in the following equation:
  
 \begin{multline}
 I(q) = G_1 \exp\left(-\frac{q^2R_{\text{g},1}^2}{3}\right)+B_1\exp\left(-\frac{q^2R_{\text{g},2}^2}{3}\right){q_1^\star}^{p_1}
\\+G_2\exp\left(-\frac{q^2R_{\text{g},2}^2}{3}\right)+B_2 {q_2^\star}^{p_2}
\label{eq:Beaucage}
\end{multline}
where
 \begin{equation}
q^\star_i=\frac{q}{\left\{\text{erf}\left(\frac{qR_{\text{g},i}}{\sqrt{6}}\right)\right\}^3}
 \end{equation}
and $i=1$ and $i=2$ respectively correspond to the large-scale and to the small-scale level, i.e. $R_{\text{g},1}\gg R_{\text{g},2}$.

 \begin{figure*}
 	\centering
 	\includegraphics[width=0.8\textwidth]{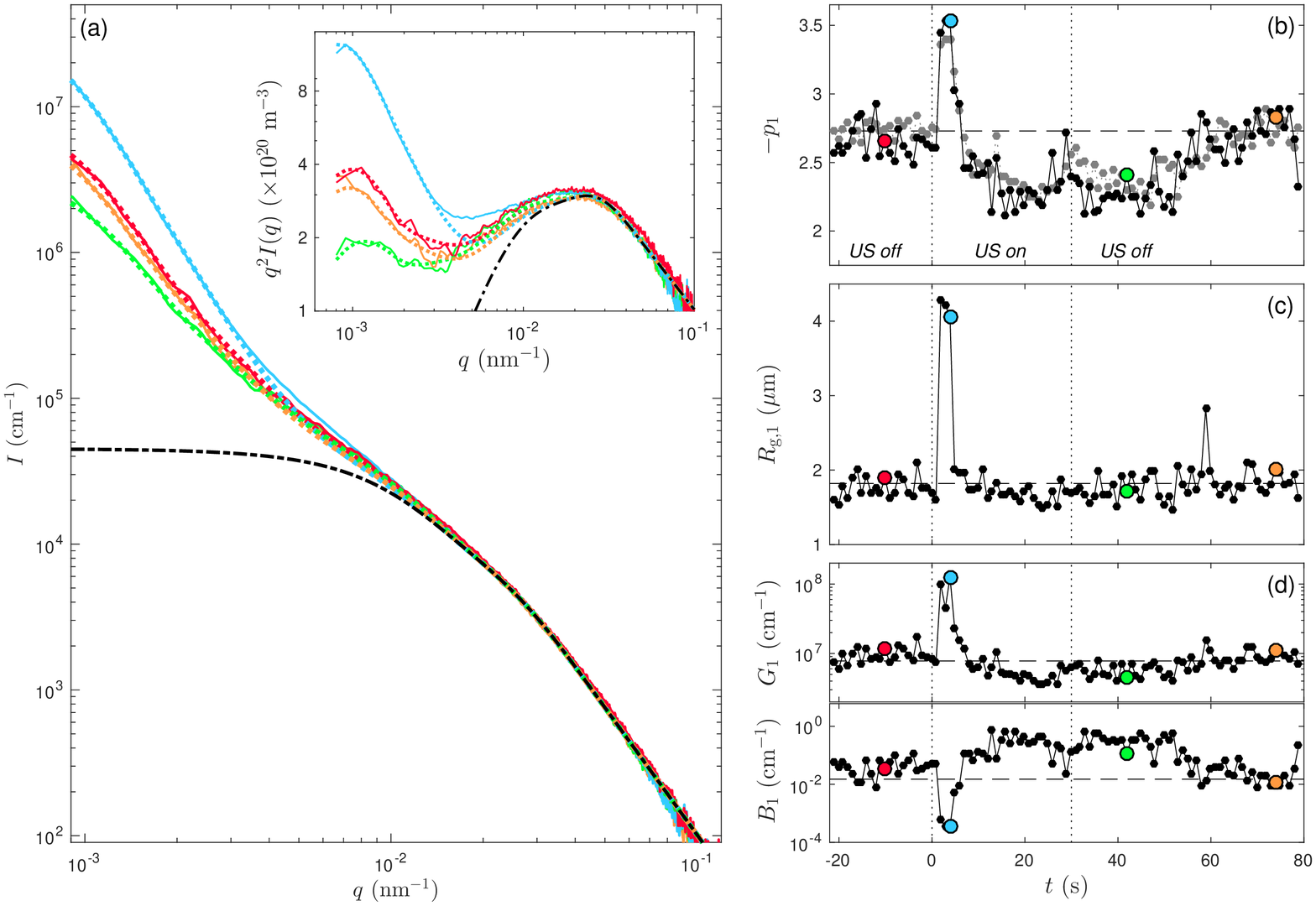}
     \caption{(a) Time-resolved USAXS intensity spectra $I(q)$ recorded in a 3~\% vol. carbon black gel under ultrasonic vibrations with frequency 20~kHz for $P=64$~kPa. Ultrasonic vibrations are turned on at $t=0$ and switched off at $t=30$~s. The times at which the spectra are shown correspond to the various colored symbols in (b,c,d). The inset shows the same data in a Kratky plot, i.e. by plotting $q^2 I(q)$ vs $q$. The solid lines are the experimental spectra while the dotted curves are the fits with the two-level Beaucage model given by Eq.~\eqref{eq:Beaucage}. The black dash-dotted curve corresponds to the form factor $I_2(q)$ discussed in the text. The four free parameters of fits to Eq.~\eqref{eq:Beaucage} are: (b)~the exponent $-p_1$, (c)~the characteristic size $R_{\text{g},1}$ and (d)~the prefactors $G_1$ and $B_1$. In (b), the fit results for $-p_1$ (in black) are compared to the power-law exponents $p$ at low $q$ (in grey) replotted from Fig.~\ref{fig:saxs}(b). See also Video~\ref{fig:movie1}.}
     \label{fig:saxs_fits_13pc}
 \end{figure*}

\begin{figure*}
 	\centering
 	\includegraphics[width=0.8\textwidth]{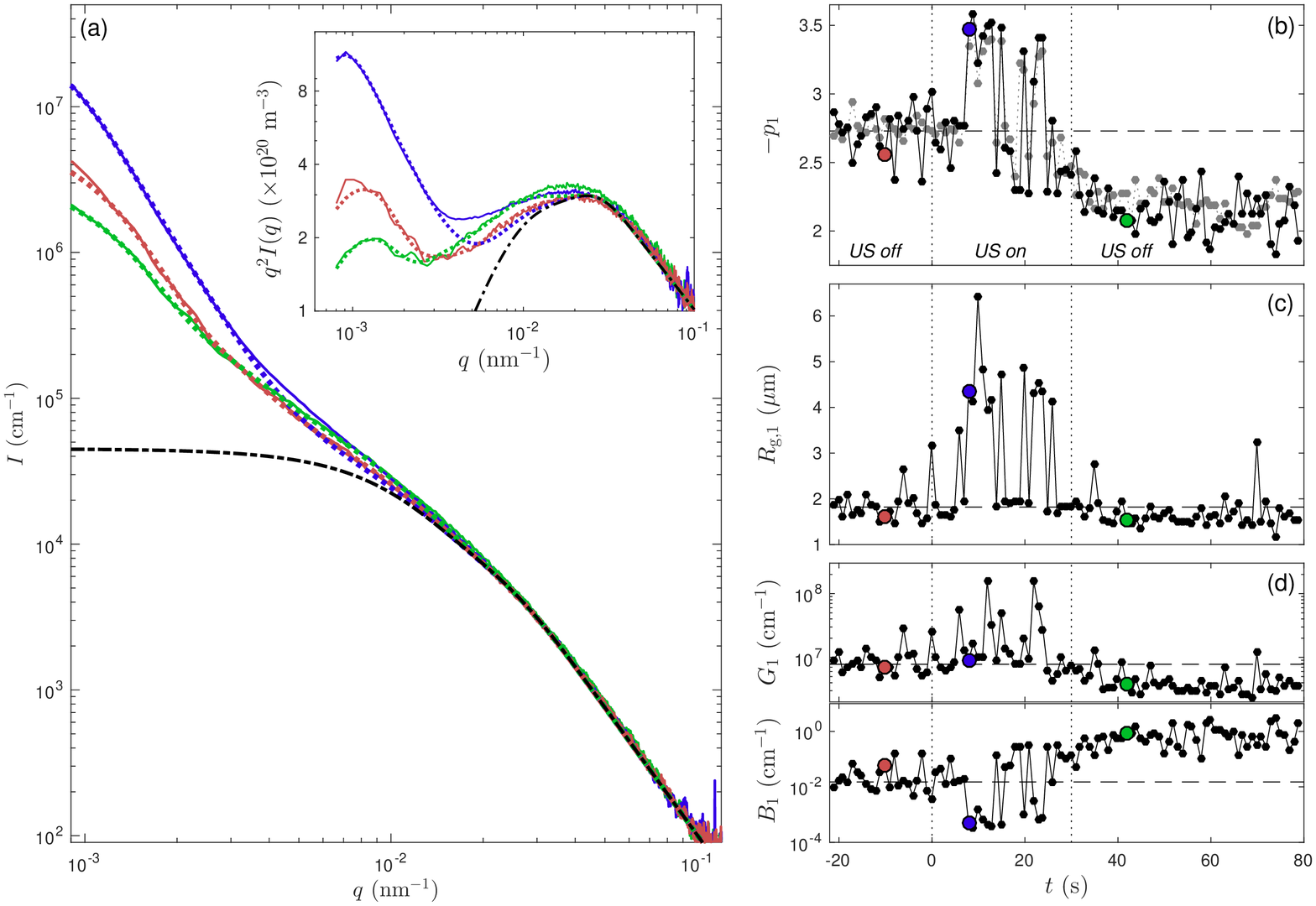}
     \caption{(a) Time-resolved USAXS intensity spectra $I(q)$ recorded in a 3~\% vol. carbon black gel under ultrasonic vibrations with frequency 20~kHz for $P=240$~kPa. Ultrasonic vibrations are turned on at $t=0$ and switched off at $t=30$~s. The times at which the spectra are shown correspond to the various colored symbols in (b,c,d). The inset shows the same data in a Kratky plot, i.e. by plotting $q^2 I(q)$ vs $q$. The solid lines are the experimental spectra while the dotted curves are the fits with the two-level Beaucage model given by Eq.~\eqref{eq:Beaucage}. The black dash-dotted curve corresponds to the form factor $I_2(q)$ discussed in the text. The four free parameters of fits to Eq.~\eqref{eq:Beaucage} are: (b)~the exponent $-p_1$, (c)~the characteristic size $R_{\text{g},1}$ and (d)~the prefactors $G_1$ and $B_1$. In (b), the fit results for $-p_1$ (in black) are compared to the power-law exponents $p$ at low $q$ (in grey) replotted from Fig.~\ref{fig:saxs}(c). See also  Video~\ref{fig:movie2}.}
     \label{fig:saxs_fits_100pc}
 \end{figure*}

 We first fix the small-scale level by fitting the high-$q$ range of the time-averaged spectrum measured in the absence of ultrasound by a single unified function: $I_2(q) = G_2\exp(-q^2R_{\text{g},2}^2/3)+B_2 {q_2^\star}^{p_2}$. We find that $R_{\text{g},2}=150$~nm, $p_2=-2.9$, $G_2=4.5\,10^4$~cm$^{-1}$ and $B_2=0.128$~cm$^{-1}$ provide a very good description of $I(q)$ for $q>10^{-2}$~nm$^{-1}$ [see black dash-dotted curves in Figs.~\ref{fig:saxs_fits_13pc}(a) and \ref{fig:saxs_fits_100pc}(a)]. The characteristic size of 150~nm is fully consistent with both the radius of gyration (177~nm) and the hydrodynamic radius (135~nm) reported recently for the same carbon black particles based on neutron and light scattering \cite{Richards:2017}. The exponent $-2.9$ also agrees well with the fractal dimension of 2.7 measured on these particles by the same authors. It is observed in Figs.~\ref{fig:saxs_fits_13pc}(a) and \ref{fig:saxs_fits_100pc}(a) that the high-$q$ region of the USAXS spectra remains unaffected by ultrasonic vibrations whatever the acoustic pressure $P$. This means that ultrasound does not modify the structure at the scale of the carbon black particles, which is consistent with the fact that these particles are themselves unbreakable aggregates of permanently fused primary particles of typical diameter 20~nm. Thus, the small-scale level $I_2(q)$ can be taken as the form factor of the carbon black particles and remains unchanged throughout all experiments.
 
 Ultrasound, however, has a strong impact on the larger scales ($q<10^{-2}$~nm$^{-1}$). To quantify this impact, we fit the USAXS spectra for $q>10^{-3}$~nm$^{-1}$ to Eq.~\eqref{eq:Beaucage} where $R_{\text{g},2}$, $p_2$, $G_2$ and $B_2$ are fixed to the previous values and $R_{\text{g},1}$, $p_1$, $G_1$ and $B_1$ are taken as free fitting parameters. Fit results are shown with dotted lines in Figs.~\ref{fig:saxs_fits_13pc}(a) and \ref{fig:saxs_fits_100pc}(a) for a few spectra representative of the time-resolved USAXS data before (red), during (blue) and just after sonication (green). The full set of fitting parameters are shown as a function of time in the right panels of Figs.~\ref{fig:saxs_fits_13pc} and \ref{fig:saxs_fits_100pc}. For both ultrasonic pressures under investigation, the evolution of the exponent $-p_1$ is found to be nicely consistent with the power-law exponent $p$ [compare black and grey bullets in Figs.~\ref{fig:saxs_fits_13pc}(b) and \ref{fig:saxs_fits_100pc}(b)]. 
 
For the lower acoustic intensity ($P=64$~kPa, see Fig.~\ref{fig:saxs_fits_13pc} and Video~\ref{fig:movie1}), all fitting parameters show a strong peak upon turning on ultrasonic vibrations. This peak coincides with that observed in the viscous modulus concomitantly with the strong decrease of the elastic modulus in Fig.~\ref{fig:modulus_normalized}(c). The fits with the two-level Beaucage model show that this peak is associated with a transient increase in the low-$q$ scattering that turns from that of a mass fractal ($-p_1<3$) to that of a surface fractal ($-p_1>3$) with a fractal dimension $6+p_1\simeq 2.5$ \cite{Schmidt:1991}. Since low enough $q$ values are not accessible to the present USAXS measurements, the crossover to the Guinier regime is not fully observed so that $R_{\text{g},1}$ can only be taken as a rough estimate of a characteristic microstructural size under ultrasonic vibrations. Still, the time evolution reported here points to the scenario illustrated in the sketches of Fig.~2 in the main text, where the initial microstructure, made of a percolated network of micronsized fractal aggregates of carbon black particles, transiently breaks up into a fragmented assembly of aggregates separated by patches of solvent with a characteristic size of several micrometers. After this initial break-up, the microstructure heals back to a space-spanning network yet with lower fractal dimension. Once ultrasonic vibrations are turned off, the system further relaxes until it fully recovers its initial microstructure over the course of about one minute [compare red and orange spectra in Fig.~\ref{fig:saxs_fits_13pc}(a)].

For the larger acoustic intensity ($P=240$~kPa, see Fig.~\ref{fig:saxs_fits_100pc} and Video~\ref{fig:movie2}), the temporal response consists of a series of peaks where the spectra change from mass fractal to surface fractal with fit parameters similar to those observed within the single peak measured at the lower intensity. This suggests that the microstructure constantly changes from a space-spanning network of clusters to a fractured network with micronsized cracks filled with solvent. At this stage, it is not clear whether this unsteadiness results from a global evolution of the whole sample or from more local effects. Indeed the spectra are averaged over the size of the x-ray beam ($80~\mu$m$\times 150~\mu$m) so that the strong fluctuations in the USAXS data could be caused by large-scale motion of carbon black clusters entering and leaving the beam. Moreover, the pressure amplitude investigated here in the USAXS setup is above the range of pressures that can be achieved with the rheometer setup, which makes it difficult to directly compare the USAXS data with the evolution of global features such as the elastic moduli. Once ultrasonic vibrations are stopped, the characteristic radius $R_{\text{g},1}$ goes back to its initial value within experimental uncertainty but the Porod contribution $B_1$ to the low-$q$ structural level remains significantly larger than the initial one, with an exponent $-p_1$ ranging between 2 and 2.5 indicative of a microstructure looser than before application of ultrasound. 

 \begin{video}
 	\centering
 	\includegraphics[width=0.3\textwidth]{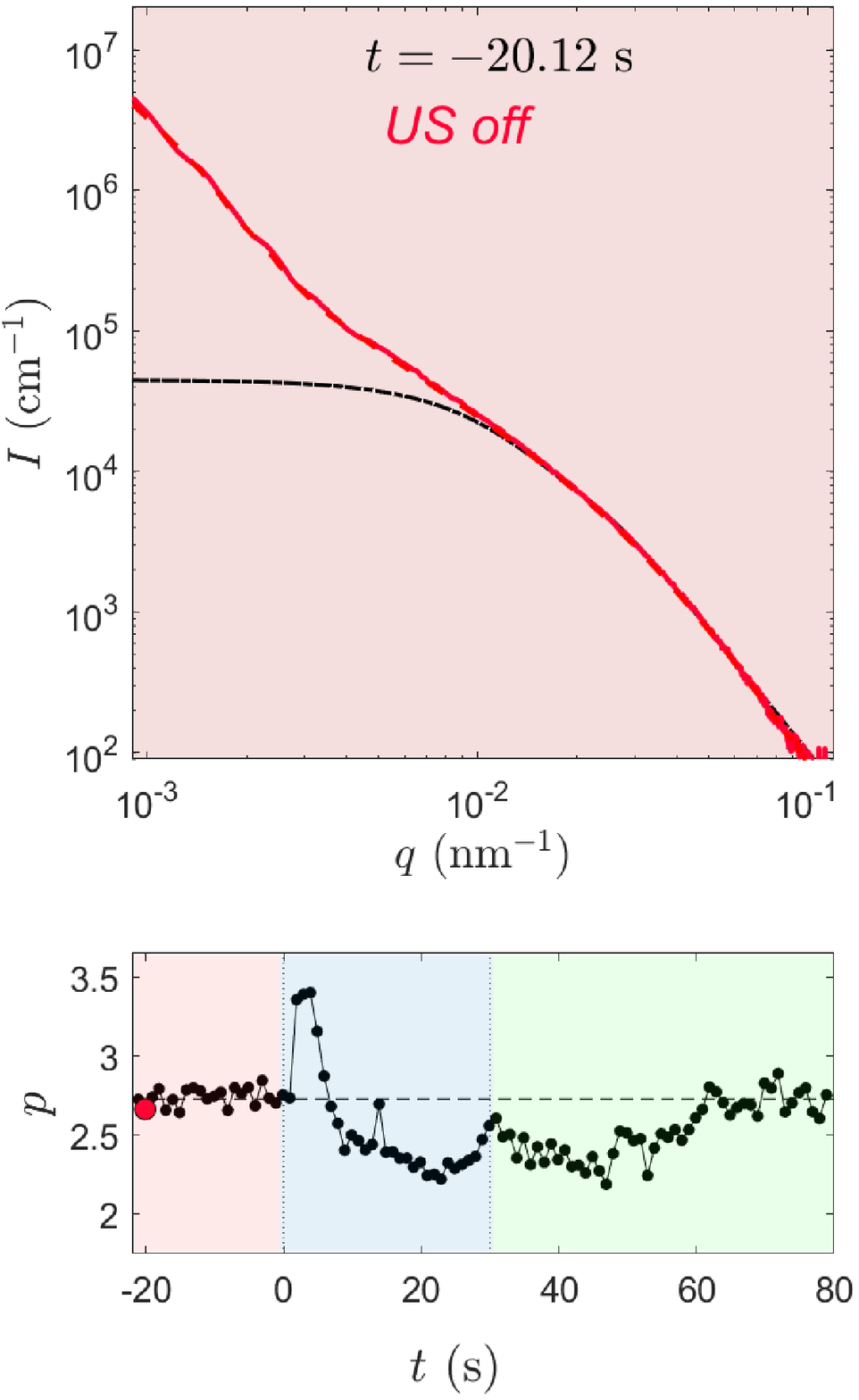}
     \caption{Time-resolved USAXS measurements on a 3~\% vol. carbon black gel under ultrasonic vibrations with frequency 20~kHz for $P=64$~kPa. The data correspond to those analyzed in Fig.~\ref{fig:saxs_fits_13pc}. The top panel shows the USAXS intensity spectra $I(q)$ as a function of time. The bottom panel displays the time evolution of the exponent $p$ obtained from a power-law fit of $I(q)\sim q^{-p}$ at low $q$. Ultrasonic vibrations are turned on at $t=0$ and switched off at $t=30$~s. The black dash-dotted curve corresponds to the form factor $I_2(q)$ discussed above. The red dashed line shows the initial spectrum prior to application of ultrasound. The current measurement is drawn in red, blue or green when recorded respectively before, during or after application of ultrasound.}
     \label{fig:movie1}
 \end{video}
 
  \begin{video}
 	\centering
 	\includegraphics[width=0.3\textwidth]{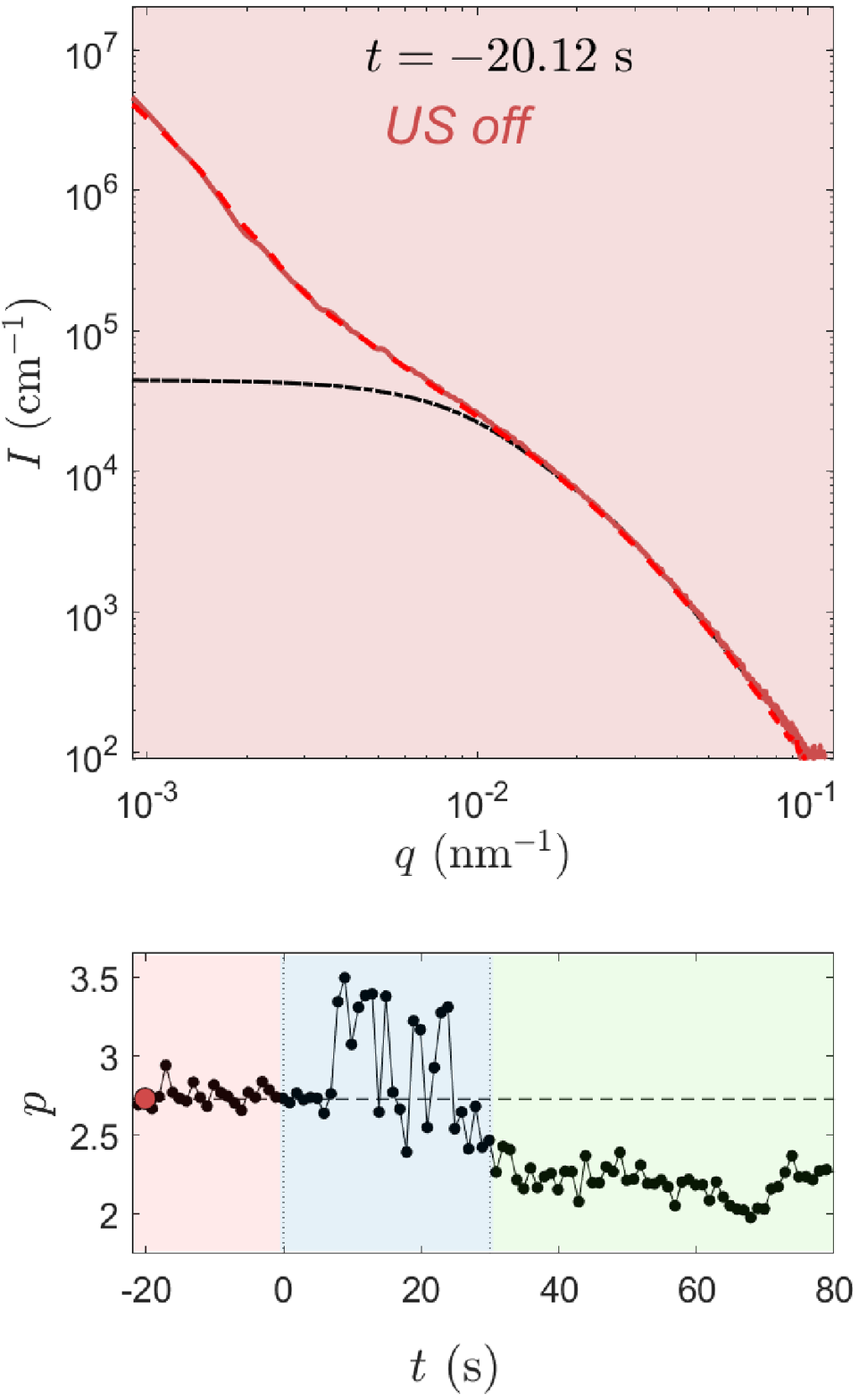}
     \caption{Same as Video~\ref{fig:movie1} for $P=240$~kPa. The data correspond to those analyzed in Fig.~\ref{fig:saxs_fits_100pc}.}
     \label{fig:movie2}
 \end{video}

 \begin{figure}[h!]
 	\centering
 	\includegraphics[width=0.45\textwidth]{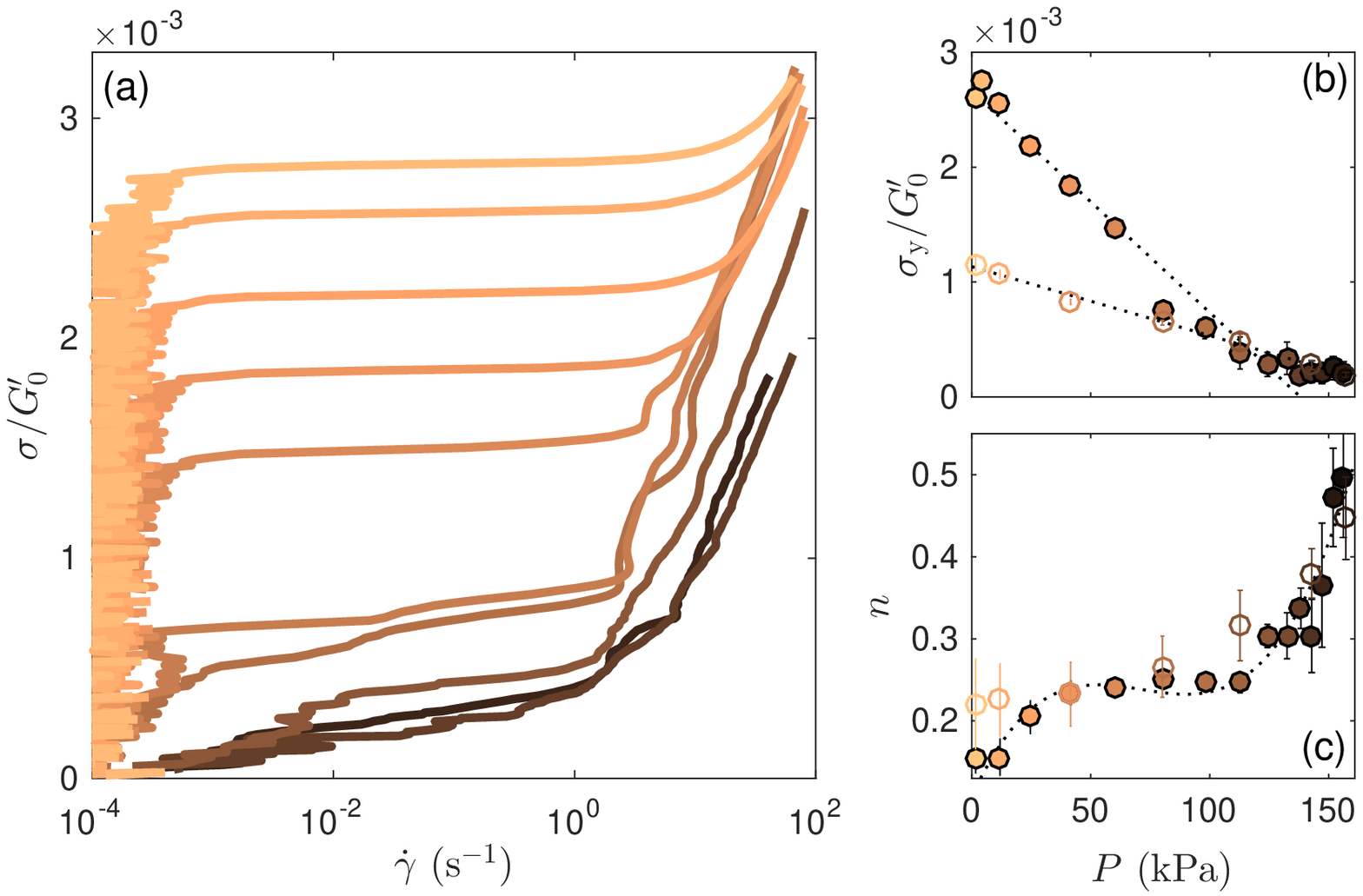}
     \caption{(a)~Flow curves, shear stress $\sigma$ vs shear rate $\dot\gamma$, measured in a 10~\% vol. calcite gel for different acoustic intensities [increasing from top to bottom, see color code in (b,c)]. The frequency of ultrasonic vibrations is 45~kHz. The shear stress is normalized by the elastic modulus $G'_0$ measured at rest. (b)~Normalized yield stress $\sigma_\text{y}/G'_0$ and (c)~shear-thinning exponent $n$ as a function of the acoustic pressure $P$. In (b) and (c), the results obtained from the flow curves shown in (a) and measured by sweeping up the shear stress are plotted with filled symbols and compared to the results of Fig.~\ref{fig:gdot_down}(b) and (c) (open symbols) obtained from downward sweeps of the shear rate as described in Appendix~\ref{app:protocol_rheo}.}
     \label{fig:stress_up}
 \end{figure}
 
\subsection{Flow properties under ultrasonic vibrations: complementary flow curve measurements}
\label{app:stress_up}
 
To complement the flow curve measurements of Fig.~3 that characterized the dynamic yield stress in the main text, Fig.~\ref{fig:stress_up} shows the results of increasing stress ramps in the calcite gel. This protocol allows us to define the static yield stress as the stress at which the shear rate suddenly jumps from a vanishingly small value ($\dot\gamma<10^{-3}$~s$^{-1}$) to a finite value ($\dot\gamma>0.3$~s$^{-1}$). Error bars on the static yield stress are estimated by taking the standard deviation of the stress measured over the plateau that separates those shear rate values. As for the dynamic yield stress, the static $\sigma_\text{y}/G_0$ decreases linearly with the acoustic pressure $P$ [Fig.~\ref{fig:stress_up}(b)]. Interestingly, the normalized static yield stress remains more than two times larger than its dynamic counterpart as long as $P\lesssim 70$~kPa while the two yield stresses become indistinguishable at larger acoustic pressures. For $P\gtrsim 70$~kPa, all flow curves are monotonically increasing and the shear-thinning index $n$ strongly increases with $P$ [Fig.~\ref{fig:stress_up}(c)] as already reported in Fig.~3(c). Here, error bars on $n$ correspond to the dispersion obtained on the exponent of power-law fits at high shear rate by varying the fitting interval. 

Below the transition at $P\simeq 70$~kPa, the flow curve displays a decreasing part under an imposed shear rate [Fig.~3(a)] and a true stress plateau under an imposed stress [Fig.~\ref{fig:stress_up}(a)]. These two features are characteristic of attractive systems where flow competes with restructuring dynamics, leading to mechanical instability and shear banding \cite{Bonn:2017}. On the other hand, flow curves that increase monotonically above the yield point are typical of repulsive soft glassy materials. Therefore, we may  conclude that for the present calcite gel, ultrasonic vibrations induce a transition from an attraction-dominated system to a repulsive-like yield stress fluid with weaker and weaker shear-thinning properties as the acoustic intensity is increased.

\subsection{Ultrasound-assisted fluidization: accounting for ultrasound through an effective temperature}
\label{app:Teff}
 
In the main text, the master curve in Fig.~4(f) shows that, for the carbon black gel involved in the present study, the fluidization time $\tau_\text{f}$ under a stress $\sigma$ and an acoustic pressure $P$ can be written as:
\begin{equation}
    \tau_\text{f}(\sigma,P)=\tilde{\tau}\exp\left(\frac{\tilde{\sigma}-\sigma}{\sigma_0(P)}\right)\,,
\label{eq:fluidization_data}
\end{equation}
where $\sigma_0(P)$ is displayed in Fig.~4(c) and the parameters $\tilde{\tau}=57.1$~s and $\tilde{\sigma}=24.9$~Pa are inferred from Fig.~4(e). We introduce the {\it effective} temperature defined as:
\begin{equation}
   T_\text{eff}(P)=\frac{\sigma_0(P)}{\sigma_0(0)}\,T\,,
\label{eq:Teff}
\end{equation}
such that the effective temperature coincides with the thermodynamic temperature in the absence of ultrasonic vibrations, i.e. for $P=0$. This allows us to recast Eq.~\eqref{eq:fluidization_data} in terms of $T_\text{eff}$ and $T$ as:
\begin{equation}
    \tau_\text{f}(\sigma,P)=\tilde{\tau}\exp\left(\frac{(\tilde{\sigma}-\sigma)T}{\sigma_0(0)T_\text{eff}(P)}\right)\,.
\label{eq:fluidization_dataTeff}
\end{equation}

In its simplest form, i.e. when taking the ``high stress limit'' where all bonds in a strand cross-section break almost simultaneously, the mean-field model introduced in Refs.~\cite{Sprakel:2011,Lindstrom:2012} predicts a fluidization time:
\begin{equation}
    \tau_\text{f}^\text{m}(\sigma)=\frac{1}{\omega_0C\sigma}\exp\left(\frac{E_\text{A}}{k_\text{B}T}-C\sigma\right)\,,
\label{eq:fluidization_model}
\end{equation}
where $\omega_0$ is the attempt frequency for bond dissociation, $C=\delta/(n\rho_0 k_\text{B}T)$ with $n$ the average number of bonds per cross-section of a network strand and, as defined in the main text, $\rho_0$ the initial area density of particles, $\delta$ the range of the interaction potential and $E_\text{A}$ the depth of the potential well. $C$ thus corresponds to the inverse of the characteristic stress $\sigma_0$ in the absence of ultrasonic vibrations. Assuming that the effect of ultrasonic vibrations with amplitude $P$ can be captured through an effective temperature $T_\text{eff}(P)$ and neglecting the logarithmic correction $1/C\sigma$, we may rewrite the model prediction in the presence of ultrasonic vibrations as:
\begin{equation}
    \tau_\text{f}^\text{m}(\sigma,P)=\tilde{\tau}^\text{m}\exp\left(\frac{E_\text{A}}{k_\text{B}T_\text{eff}(P)}-\frac{\delta\sigma}{n\rho_0 k_\text{B}T_\text{eff}(P)}\right)\,.
\label{eq:fluidization_modelTeff}
\end{equation}

Identification of Eq.~\eqref{eq:fluidization_dataTeff} and Eq.~\eqref{eq:fluidization_modelTeff} for any $\sigma$ and $P$ leads to:
\begin{equation}
\frac{E_\text{A}}{k_\text{B}T}=\frac{\tilde{\sigma}}{\sigma_0(0)}.
\label{eq:depth}
\end{equation}
With the experimentally measured values $\tilde{\sigma}=24.9$~Pa and $\sigma_0(0)=1.2$~Pa, Eq.~\eqref{eq:depth} yields $E_\text{A}\simeq 20.8\,k_\text{B}T$. This estimate of the interaction energy nicely falls into the range of values previously published for carbon black, typically 10--$30\,k_\text{B}T$, depending on the solvent and on the presence of dispersant \cite{Trappe:2001,Prasad:2003,Trappe:2007}. Together with the results shown in Fig.~\ref{fig:fluidization}, this strongly supports a mean-field interpretation of ultrasound-assisted fluidization in the framework of effective temperatures.

\subsection{Influence of the acoustic frequency}
\label{app:frequency}
 
 \begin{figure}[htb]
 	\centering
 	\includegraphics[width=0.35\textwidth]{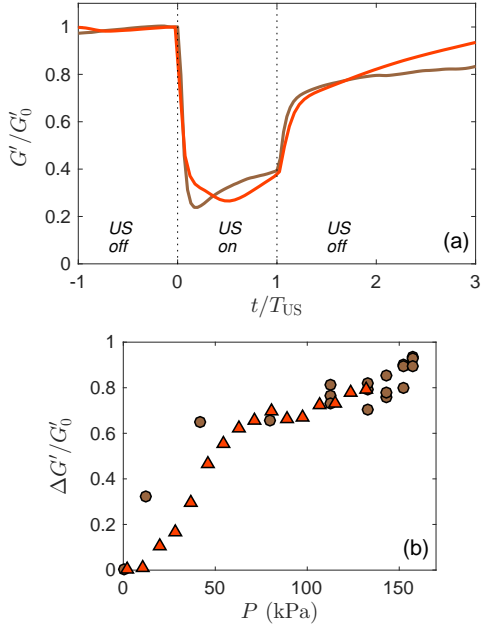}
     \caption{(a)~Elastic modulus $G'$ as a function of time $t$ for a 10~\% vol. calcite gel. Ultrasonic vibrations are applied at $t=0$ for a duration $T_\text{US}$ with a frequency of 45~kHz (brown curve, $T_\text{US}=30$~s, $P=113$~kPa) and 500~kHz (red curve, $T_\text{US}=20$~s, $P=115$~kPa) at about the same acoustic pressure $P$. The elastic modulus is normalized by its value $G'_0$ measured just before ultrasonic vibrations are turned on and time is normalized by $T_\text{US}$. (b)~Relative amplitude of the softening effect $\Delta G'/G'_0$ as a function of $P$ at 45~kHz (brown bullets) and 500~kHz (red triangles). The upper plate is covered with sandpaper. Experiments at 45~kHz were repeated three times for $P\geq 112$~kPa in order to check for reproducibility.}
     \label{fig:modulus_frequency}
 \end{figure}
 
 \begin{figure}[htb]
 	\centering
 	\includegraphics[width=0.4\textwidth]{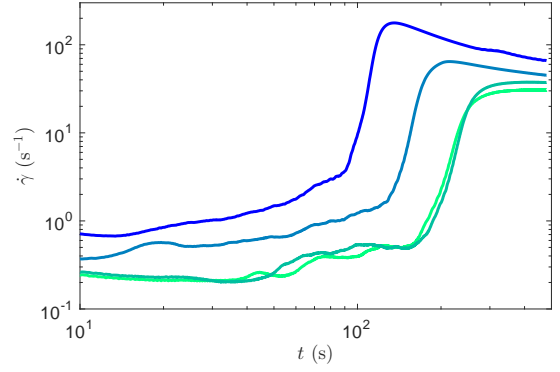}
     \caption{Shear rate responses $\dot\gamma(t)$ measured in a 3~\% vol. carbon black gel under a constant shear stress $\sigma=22$~Pa. The two lower curves (in green) are obtained in the absence of ultrasonic vibrations, respectively before and after the two upper curves, in order to illustrate reproducibility. The two blue curves are measured under ultrasound with frequency 500~kHz and pressure amplitude $P=55$~kPa (middle curve) and $P=108$~kPa (top curve). In order to avoid damaging the transducer, ultrasonic vibrations are turned on for only 200~s and 120~s respectively. The upper plate is sandblasted and the gap size is 0.5~mm.}
     \label{fig:fluidization_frequency}
 \end{figure}
 
Figure~\ref{fig:modulus_frequency} shows that for the same pressure amplitude $P$, the effects of ultrasonic vibrations on the elastic modulus of the calcite gel almost superimpose at 45~kHz and at 500~kHz [Fig.~\ref{fig:modulus_frequency}(a)]. In particular, the relative softening amplitudes $\Delta G'/G'_0$ follow the same evolution and even coincide to within experimental dispersion for $P>70$~kPa [Fig.~\ref{fig:modulus_frequency}(b)]. Therefore, a ten-fold increase in the acoustic frequency does not affect the softening effect.

The same conclusion is reached for ultrasound-assisted fluidization. Figure~\ref{fig:fluidization_frequency} shows delayed yielding experiments on the carbon black gel under ultrasonic vibrations with frequency 500~kHz. As already observed in the main text in Fig.~4 for an ultrasonic frequency of 45~kHz, ultrasonic vibrations strongly accelerate delayed yielding and stress-induced fluidization. Technical limitations of the 500-kHz transducer, which cannot be used continuously for more than one minute at the highest intensities, do not allow for a quantitative comparison. Still, the effects of ultrasonic vibrations on the fluidization time are very similar for both frequencies.

 \subsection{Influence of the geometry gap}
 \label{app:geometry}
 
  \begin{figure}[htb]
 	\centering
 	\includegraphics[width=0.4\textwidth]{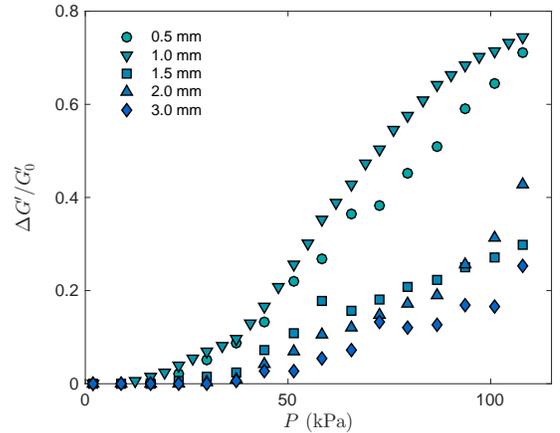}
     \caption{Relative amplitude of the softening effect $\Delta G'/G'_0$ as a function of $P$ in a 3~\% vol. carbon black gel for different gap sizes as indicated in the legend. The upper plate is sandblasted and the frequency of ultrasonic vibrations is 500~kHz.}
     \label{fig:modulus_geometry}
 \end{figure}
 
 All data presented in the main text were obtained with the same gap size of 1~mm separating the ultrasonic transducer and the upper plate attached to the rheometer. Figure~\ref{fig:modulus_geometry} explores the softening effect discussed in Fig.~1 as a function of the gap size in the carbon black gel under ultrasonic vibrations at 500~kHz. While gaps of widths 0.5 and 1~mm yield very similar values of $\Delta G'/G'_0$, the softening is about two times weaker in gaps of widths 1.5, 2 and 3~mm. Such a difference could be due to variations in the structure of the ultrasonic field within the parallel-plate device, depending on reflections on the upper plates and on interference conditions at the transducer surface. Indeed, the wavelength at 500~kHz in carbon black is about 3~mm, which is comparable to the larger gap sizes. A more thorough investigation of the effects of the geometry is left for future work. Overall, however, ultrasound-induced softening is robustly observed for all gaps under study.

\subsection{Influence of a temperature ramp}
\label{app:temperature}

 \begin{figure}[htb]
 	\centering
 	\includegraphics[width=0.35\textwidth]{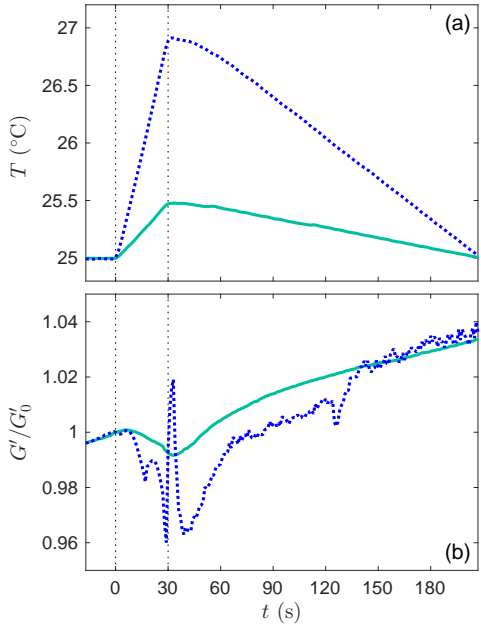}
     \caption{(a)~Temperature $T$ imposed to a 3~\% vol. carbon black gel as a function of time $t$ for two different ramps: starting from $T=25^\circ$C, the temperature is increased linearly by $0.5^\circ$C (solid line) or by $2^\circ$C (dotted line) within 30~s. $T$ is then decreased back to $25^\circ$C over 180~s. (b)~Elastic modulus $G'$ normalized by its value $G'_0$ measured at the start of the temperature increase and plotted as a function of time during the ramps shown in (a) using corresponding lines and colors. Experiments conducted in a parallel-plate geometry of gap 1~mm with a bottom plate equipped with a Peltier element and a sandblasted upper plate.}
     \label{fig:temperature}
 \end{figure}

At the highest achievable acoustic powers, the ultrasonic transducers used in this study dissipate a significant amount of heat. Although the elastic modulus of carbon black gels at rest is known to increase with temperature \cite{Won:2005}, we need to check that a fast, transient heating of the gel cannot account for the softening observed under ultrasonic vibrations. To this aim, the carbon black gel was submitted to a temperature ramp in a parallel-plate geometry thanks to the standard bottom plate of the rheometer equipped with a Peltier element. The parameters of the two temperature ramps shown in Fig.~\ref{fig:temperature}(a) are chosen so that they mimic the temperature increases induced by ultrasonic vibrations during 30~s, respectively at high intensities (1$^\circ$C\,min$^{-1}$ i.e. $P\simeq 100$~kPa) and very high intensities (4$^\circ$C\,min$^{-1}$ i.e. $P\simeq 150$~kPa). For the slowest ramp, the elastic modulus decreases by less than 1~\% [see light blue solid curve in Fig.~\ref{fig:temperature}(b)]. For the fastest ramp, the evolution of $G'/G'_0$ is more complex, most probably due to the dilation of the bottom plate, but the effect on the elastic modulus is at most 4~\%, well below the softening amplitudes of 40\%--90~\% reported under ultrasonic vibrations at the highest acoustic intensities. This allows us to rule out any significant effect of the heating due to dissipation within the ultrasonic transducers on ultrasound-induced softening.

\subsection{Control experiments}
\label{app:control}

\begin{figure}[b]
 	\centering
 	\includegraphics[width=0.35\textwidth]{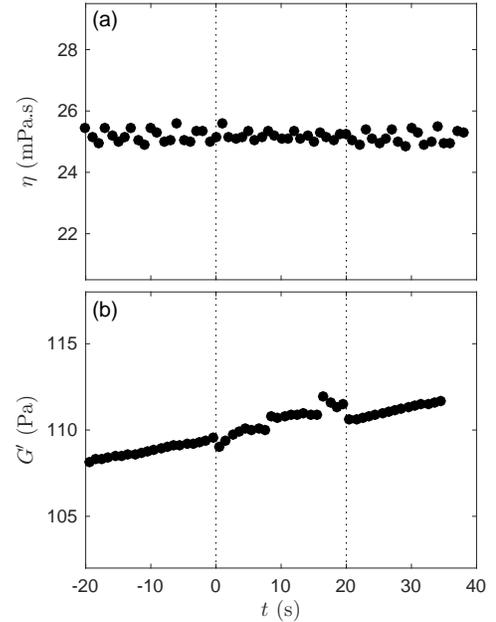}
     \caption{(a)~Viscosity $\eta$ of a Newtonian oil (light mineral oil, Sigma-Aldrich) and (b)~elastic modulus $G'$ of a 2~\% wt. carbopol microgel as a function of time $t$. Ultrasonic vibrations with frequency 45~kHz and pressure amplitude $P=110$~kPa are turned on at time $t=0$ and switched off at $t=20$~s. Experiments conducted in a cone-and-plate geometry of angle 1$^\circ$, diameter 25~mm and truncation of 47~$\mu$m. In (a), the viscosity is measured under an applied shear rate of 100~s$^{-1}$. In (b), the elastic modulus is measured under an oscillatory strain of amplitude 1~\% and frequency 1~Hz.}
     \label{fig:control}
 \end{figure}
 
In order to check that ultrasonic vibrations do not interfere with rheological measurements, we performed control experiments in a Newtonian fluid (a light mineral oil) and in a simple yield stress fluid (a 2~\% wt. carbopol microgel prepared as described in Ref.~\cite{Lidon:2017}). Figure~\ref{fig:control}(a) shows that viscosity measurements in the Newtonian fluid are unaffected by the application of large-amplitude ultrasonic vibrations at 45~kHz. Similarly, as seen in Fig.~\ref{fig:control}(b), the elastic modulus $G'$ of the microgel does not change by more than 2~\% and does not show any systematic trend when ultrasonic vibrations are applied with the same frequency and pressure amplitude. This is expected as the carbopol microgel is constituted of a dense assembly of soft, swollen polymer particles: contrary to the fragile structure of the colloidal gels investigated in the main text, the microstructure of this soft jammed material appears to be insensitive to ultrasound. This confirms that neither the torque transducer nor the optical strain gauge of our stress-imposed rheometer are perturbed by ultrasonic vibrations applied at the bottom plate of the geometry.

\end{document}